\documentclass[onecolumn,nofootinbib,longbibliography,preprintnumbers,amsmath,amssymb,authblk,10pt]{revtex4-2}

\usepackage{amsmath,amssymb, tabularx,booktabs,physics,verbatim,soul,xcolor,tikz}

\usetikzlibrary{shapes, positioning, calc, arrows}

\usepackage{amsthm}

\theoremstyle{definition}
\newtheorem{definition}{Definition}[section]
\newtheorem{fact}{Fact}
\newtheorem{assumption}{Assumption}
\newtheorem{theorem}{Theorem}
\newtheorem{lem}{Lemma}

\newtheorem{cor}{Corollary}

\newcommand{\GenP}[3]{\ensuremath{\hat{#1}^{{#2}_{#3}}}}
\newcommand{\GGenP}[3]{\ensuremath{\GenP{#1}{#3}{1}\GenP{#2}{#3}{2}}}
\newcommand{\compop}[5]{\ensuremath{\hat{#1}_{#2}\otimes (\hat{#3}_{#4})^{#5}}}

\newcommand{\s}[1]{\ensuremath{\hat{\sigma}_{#1}}}
\newcommand{\p}[1]{\ensuremath{\hat{P}_{#1}}}

\newcommand{\drawCirclePair}[7]{
	
	\draw (#1,#2) circle (#4);
	\node at (#1,#2) {#5};
	
	\draw (#1+#3,#2) circle (#4);
	\node at (#1+#3,#2) {#6}; 
	
	\draw[#7] (#1+#4, #2) -- (#1+#3-#4, #2); 
}

\newcommand{\drawCircle}[4]{
	
	\draw (#1,#2) circle (#4);
	\node at (#1,#2) {#3};
}

\begin{document}
	\title{The Structure of Quantum Questions}
	\author{Yang Yu}	
	\author{Philip Goyal}	
	\affiliation{University at Albany~(SUNY), NY, USA}
	\begin{abstract}

		In classical physics, a single measurement can in principle reveal the state of a system. However, quantum theory permits numerous non-equivalent measurements on a physical system, each providing only limited information about the state. This set of various measurements on a quantum system indicates a rich internal structure. We illuminate this structure for both individual and composite systems by conceptualizing measurements as questions with a finite number of outcomes. We create a mathematical question structure to explore the underlying properties, employing the concept of information as a key tool representing our knowledge gained from asking these questions. We subsequently propose informational assumptions based on properties observed from measurements on qubits, generalizing these to higher dimensional systems.
		
		Our informational assumptions shape the correlations between subsystems, which are symbolized as classical logical gates. Interestingly, systems with prime number dimensions exhibit unique property: the logical gate can be expressed simply as a linear equation under modular arithmetic. We also identify structures in quantum theory that correspond to those in the structure of quantum questions. For instance, the questions determining the system correspond to generalized Pauli matrices, and the logical gate connecting questions in subsystems is directly related to the tensor product combining operators. Based on these correspondences, we present two equivalent scenarios regarding the evolution of systems and the change of information within both quantum questions and quantum mechanics.

	\end{abstract}

	\maketitle

	\section{Introduction}
		
	A fundamental difference between classical and quantum systems is the following:~whereas a \emph{single} measurement can be performed  on a classical system which reveals the state of the system, many different and inequivalent measurements can be performed on a single qubit, each of which generally provides limited information about the state of the system. The set of possible measurements that can be performed on a quantum system has a rich internal structure.  
	
	The motivation for the present work comes from some facts observed on spin-$\frac{1}{2}$ particles. In quantum tomography, the state of a single spin-$\frac{1}{2}$ particle can be determined by probability of the Stern-Gerlach measurements. The density matrix of a single spin-$\frac{1}{2}$ particle can then be represented as:
	\begin{equation}
		\label{singleQubitDensityMatrix}
		\hat{\rho} = \frac{1}{2}(\hat{I}+\vec{r}\cdot \hat{\vec{\sigma}})
	\end{equation}
	Moreover, consider a two-body system composed of spin-$\frac{1}{2}$ system~$A$ and~$B$, the state of this composite system can be determined by a set of local measurements and global measurements\cite{twoqubitsDM}:
	\begin{equation}
		\label{twoQubitsDensityMatrix}
		\hat{\rho}_{AB} = \frac{1}{4}(\hat{I}\otimes\hat{I}+\vec{r}_A\cdot\hat{\vec{\sigma}}^A\otimes\hat{I}+\hat{I}\otimes\vec{r}_B\cdot\hat{\vec{\sigma}}^B+\sum_{i,j}\beta_{ij}\hat{\sigma}^A_i\otimes\hat{\sigma}^B_j ),
	\end{equation}
	where~$\hat{\vec{\sigma}}^A,\hat{\vec{\sigma}}^B$ are Bloch vectors on the single spin-$\frac{1}{2}$ system A and B respectively, and~$\beta_{ij}$ are real numbers.
	Two interesting properties are observed in qubits systems: 
	1. The outcome probability of a joint measurement of~$\hat{\sigma}^A_i\otimes\hat{\sigma}^B_j$ can be obtained by the combination of statistics of local measurement outcomes of~$\hat{\vec{\sigma}}^A\otimes\hat{I}$ and~$\hat{I}\otimes\hat{\vec{\sigma}}^B$, yet the single time measurement behavior of the joint measurement and local measurement are totally different.
	2. For a single qubit, we could use a set of mutually complementary measurements,~$\{\hat{\sigma}_x,\hat{\sigma}_y,\hat{\sigma}_z\}$, to do the state tomography. For two qubits, we could also do the state tomography via those locally mutually complementary measurements and the joint measurements. However, those joint measurements may not be mutually complementary; some joint measurements are even mutually commuting to each other.
	The coincidence of commutativity and complementarity suggests there may be a deeper reason.
	We tend to use information theory to explore an explanation by introducing a new structure, \textit{quantum question structure}. 
	In one hand it is possible to provide a new viewpoint of understanding quantum mechanics, on the other hand we also find some new results of quantum mechanics.
	
	
	
	
	Employing information theory into quantum mechanics is not a new thing.
	Rovelli\cite{Rovelli1996} first proposed this idea within a new interpretation of finite dimensional quantum mechanics. Since every quantum measurement can be decomposed into many projections, the whole physical system can be regarded as a collection of binary outcome measurements. Each projection is a binary question. The state of the system is the collection of all the results of these binary questions.
	It is assumed that we may obtain 1 unit of information when obtaining the result of one question. The total information we can obtain about the system is assumed to be finite. Yet the detailed structure of questions is not mentioned.
	
	Brukner and Zellinger\cite{BruknerZeilinger2002, BruknerZeilinger2009} proposed very similar assumptions of information as Rovelli's, and information is used to build the structure of measurements of quantum mechanics.
	They suggested that the information of a system is the sum of information of all complementary questions.
	In the case of a spin-$\frac{1}{2}$ particle, this assumption is very plausible. Since all complementary questions for a spin-$\frac{1}{2}$ particle are just the spin operators.
	Yet in the case of a two-body spin-$\frac{1}{2}$ particle, the definition of complementary questions is not clear. The complementary questions for single body system seems to be related to mutually unbiased operators while using mutually unbiased operators only may not be complete to describe a composite system.

	H\"{o}hn\cite{Hohn2017,Hohn2017toolbox} proposed a new reconstruction of quantum mechanics. Rovelli's assumptions are included in this reconstruction. The relation between the joint measurements and local measurements is described by a logical gate. The information of the system is defined in a new way, it is the sum of a collection of finite questions. Together with some other fundamental assumptions, it is derived that the single system's information is determined by all complementary questions. This reconstruction claimed to recover qubit quantum mechanics and some important results are derived. However, generalizing to higher dimensions is not straightforward. It appears that the allowed logical gate, derived from specific rules, has only one form and is associative in a two-dimensional case. While in higher dimensions, there exist multiple possible formations of the logical gate, among which only a few are associative.

	The above discussions are all based on qubit quantum mechanics. Some results are really impressive, yet there is a small concern: qubit quantum mechanics may not be easily extended to higher-order dimensional cases.
	We may say that an arbitrary even number dimensional quantum system can be decomposed as a many-qubit system, but how about odd number dimensional cases, say three dimensions? Indeed, one may still possibly use qubits to represent odd number dimensional systems, yet this will definitely lead to some redundancies.
	Moreover, some results may be just a coincidence of the two dimensions. For example, one core task of the reconstruction of qubit quantum mechanics is to recover/derive the structure of Pauli matrices, since the Pauli matrices are good enough to describe a qubit. However, Pauli matrices themselves are really unique; they are pairwise complementary and anti-commuting\footnote{Anti-commutativity is not widely used. For example, $\hat{\sigma}_x$ and $\hat{\sigma}_z$ are mutually complementary, but $\hat{\sigma}_x\times \hat{\sigma}_x$ and $\hat{\sigma}_z\times \hat{\sigma}_z$ commute with each other. This local complementarity and global commutativity of Pauli matrix is directly due to anti-commutativity.}. It is natural to ask what's the analogy for generalization of Pauli matrices in higher dimension.
	Moreover, we want to deal with the odd number dimensional system without any redundancy. We are curious about what happens if applying the similar formalism of information theory to higher-dimensional cases.
	
	
	
	
	Here are the main features of this article:
	\begin{enumerate}
		\item We generalize H\"{o}hn's formalism into higher dimensional cases and find several non-trivial results.
		The main difficulty of generalization emerges when dealing with composite systems. In quantum mechanics, we can use tensor product to compose measurement on more than one single system. By abstracting to perform a measurement on physical system as to ask question to a physical system, we inherit the notion of logical gate from H\"{o}hn to connect questions on single system as the analogy of tensor products. 
		In two dimensional case, there is only one choice of logical gate, the exclusive or gate. In higher dimensional case, the choice of logical gate may not be limited.
		We find a mathematical structure, \textit{orthogonal array}, to describe the classification of different logical gates. The well discussed results on orthogonal array in prime number dimensional case lead us to focus on prime number dimensional quantum mechanics only.
		
		\item We clarify the notion and definition of information, especially information of measurement and information of system.
		The definition discussed by Rovelli and Brukner and Zeilinger is not very clear.
		We restrict that the information of measurement as a function of probability distributions of the outcomes. In some sense this is a measure of uncertainty of the outcomes. Moreover, all the probabilities we deal with are written in the Bayesian style. In this way, we may show that there are two different understanding of information of measurement and in this paper we only use one of them. The information of system is proposed under the viewpoint of tomography, where a finite set of selected measurements can be used to determine the state of system. By holding the similar assumption, the combination of information of those selected measurements characterizes all our knowledge about the system and name it as information of system.
		
		\item We provide a connection between the quantum question structure we begin with and the ordinary quantum mechanics.
		In two dimensional case, all questions have binary outcomes and the system is corresponding to a qubit. Every question resembles a Pauli matrix, i.e. a Stern-Gerlach measurement on the qubit. For composite systems, every composite question resembles a joint measurement on the multi-qubit system where each joint measurement can be represented as a tensor product of Pauli matrices.
		
		For higher dimensional cases, this analogy is not clear. Indeed we may want to find an analogy of Pauli matrix in higher dimensional space such that every question is corresponding to a specific measurement on a qudit. We choose the generalized Pauli matrix that is build based on~\textit{mutually unbiased bases} (MUBs). The reason of this choice is that we think the complementarity is the most important property of Pauli matrix, which is perfectly revealed in MUBs from qubits to higher dimensional spaces.	In this way, we could translate our main results and their derivations in terms of linear space language, making the abstract formalism not so abstract.
		
	\end{enumerate}

	The goal of the current project is to elucidate the structure of these questions. 
	This paper is organized as follows. Section~\ref{section_questionsets} introduces the new question set structure. We begin with several interesting properties among qubits and the abstract structure of quantum questions, then show the whole construction of quantum questions.
	Section~\ref{section_correspondence} introduces MUBs in ordinary quantum mechanics. It is the similarity between the properties in MUBs and the consequences in quantum question structure lead to a possible connection. Such connection is discussed in Section~\ref{section_connection}, where examples are provided to illustrate the basic idea of quantum questions in terms of ordinary quantum mechanics language. The degree of freedom of the system agree with the result in ordinary quantum mechanics. 
	Two new results are stated in Appendix, as well as an illustration of the our understanding of information.
	
	
	
	
	\section{A physical system as set of questions}
	\label{section_questionsets}
	
	\subsection{Motivation of the question structure}	
	We begin our discussion by considering a qubit. There are an infinite number of projective measurements on the qubit, each of them can be represented as a unitary operator:
	
	\begin{equation}
	\hat{\sigma}_{\theta,\phi}=\ket{+_{\theta,\phi}}\bra{+_{\theta,\phi}}-\ket{-_{\theta,\phi}}\bra{-_{\theta,\phi}} \quad \theta\in [0, \pi], \phi\in [0,2\pi),
	\end{equation}
	where
	\begin{equation}
	\ket{+_{\theta,\phi}}=\frac{1}{\sqrt{2}}\begin{pmatrix}
	\cos(\frac{\theta}{2}) \\
	e^{i\phi}\sin(\frac{\theta}{2})
	\end{pmatrix}\quad\quad \ket{-_{\theta,\phi}}=\frac{1}{\sqrt{2}}\begin{pmatrix}
	\cos(\frac{\theta}{2}) \\
	-e^{-i\phi}\sin(\frac{\theta}{2})
	\end{pmatrix}.
	\end{equation}
	
	Once a projective measurement has been performed, the post-measurement state of the system will be the eigenstate of this operator, and we can infer the outcome probabilities of any other projective measurements performed on the system immediately afterwards. All these projections have the same eigenvalues, $\pm 1$. Moreover, the state of the qubit can be reconstructed via these projections.
	
	\begin{enumerate}
		\item[] \emph{Property 1.}  The state of a single qubit can be reconstructed by state tomography over any three different projective measurements.
	\end{enumerate}

	This property can be viewed as a natural consequence of the density matrix of qubit (\ref{singleQubitDensityMatrix}). Though theoretically we could choose any three different axes for the projections, three perpendicular axis are commonly chosen for the sake of calculational convenience. Moreover, the projections with perpendicular axes has another property.
	
	\begin{enumerate}
		\item[] \emph{Property 2.}  Two projective measurements, $\hat{\sigma}_{\theta,\phi}$ and $\hat{\sigma}_{\theta',\phi'}$, with perpendicular axes will be mutually unbiased\footnote{Two operators are mutually unbiased if their eigenstates or eigensubspaces are mutually unbiased, we will talk more about that in Section \ref{section_correspondence}.} to each other.
	\end{enumerate}
	
	We can choose any two direction~$\{\theta,\phi\},\{\theta',\phi'\}$ which are mutually perpendicular and the corresponding measurements are mutually unbiased. A typical example is~$\{\theta=\frac{\pi}{2},\phi=0\},\{\theta'=0,\phi'\}$, which denotes to the measurements of~$\hat{\sigma}_x,\hat{\sigma}_z$ which are mutually unbiased,
	\begin{equation}
	\left|\bra{+}\ket{0}\right|^2 = \left|\bra{-}\ket{0}\right|^2 = \left|\bra{+}\ket{1}\right|^2 = \left|\bra{-}\ket{1}\right|^2 = \frac{1}{2}.
	\end{equation}
	
	
	\emph{Property 1\&2} are two common facts of single qubit. Noticing that those two properties are describing measurements on an individual system, and we may call them local measurements. The mutually unbiasedness between local measurements are very obvious and widely used. What attracts us more is the mutually unbiasedness between global measurements on a composite system. Here, composite system denotes to a collection of identical systems and global measurements are the tensor products of local measurements. For a composite system containing several qubits, the global measurement may be in the form of~$\hat{\sigma}_{i_1}\otimes\hat{\sigma}_{i_2}\otimes\cdots\otimes\hat{\sigma}_{i_n}$ where each~$\hat{\sigma}_{i_r}$ is a local measurement on a subsystem. 
	These global measurements have the same number of outcomes with local measurement, but the relations between global measurements will be more different, they could commute or be mutually unbiased to each other.
	We may show two interesting properties about these global measurements by taking examples on a two-body qubit system.
	
	
	\begin{enumerate}
		\item[] \emph{Property 3.} In a two-body qubit system, two global measurements~$\hat{\sigma}_{i_1} \otimes \hat{\sigma}_{j_1}$ and~$\hat{\sigma}_{i_2} \otimes \hat{\sigma}_{j_2}$ either commute or be mutually unbiased, and then they are non-informative \footnote{Addressing the unclear definition of complementary questions in Brukner and Zellinger's work, we propose using the set of non-informative questions as a complete description of a composite system. This set contains both mutually unbiased operators as well as commute operators, akin to the joint terms in the density matrix of two-body spin-$\frac{1}{2}$ system.} to each other
		, where~$\hat{\sigma}_{i_1},\hat{\sigma}_{i_2},\hat{\sigma}_{j_1},\hat{\sigma}_{j_2} \in \{\hat{\sigma}_x,\hat{\sigma}_y,\hat{\sigma}_z\} $.
	\end{enumerate}

	Two measurements are non-informative, which means the result of one measurement alone cannot determine the result of another measurement if performed subsequently. We will show this in detail for global measurements that commute or mutually unbiased.
	
	\begin{itemize}
		\item $\hat{\sigma}_{i_1} \otimes \hat{\sigma}_{j_1}$ and~$\hat{\sigma}_{i_2} \otimes \hat{\sigma}_{j_2}$ commute
		
		We first take the example on~$\hat{\sigma}_{i_1}=\hat{\sigma}_{j_1}=\hat{\sigma}_x$ and~$\hat{\sigma}_{i_2}=\hat{\sigma}_{j_2}=\hat{\sigma}_z$.
		Noticing that~$\hat{\sigma}_x\otimes \hat{\sigma}_x$ and~$\hat{\sigma}_z\otimes \hat{\sigma}_z$ commute, their common eigenstates are just the famous Bell states:
		\begin{equation}
			\begin{aligned}
				\ket{\Psi^+} &= \frac{1}{\sqrt{2}} (\ket{0}\ket{0}+\ket{1}\ket{1}) \\
				\ket{\Psi^-} &= \frac{1}{\sqrt{2}} (\ket{0}\ket{0}-\ket{1}\ket{1})\\
				\ket{\Phi^+} &= \frac{1}{\sqrt{2}} (\ket{0}\ket{1}+\ket{1}\ket{0})\\
				\ket{\Phi^-} &= \frac{1}{\sqrt{2}} (\ket{0}\ket{1}-\ket{1}\ket{0}).
			\end{aligned}
		\end{equation}

		For an unknown system, if we first take measurement of~$\hat{\sigma}_x\otimes \hat{\sigma}_x$ at time~$t_1$ and get some eigenvalue~$\pm 1$, after this measurement the state of the system would be projected into one of the two eigensubspaces of~$\hat{\sigma}_x\otimes \hat{\sigma}_x$:
		\begin{equation}
			\ket{\psi}_{t>t_1} \in \begin{cases}
				Span(\{\ket{\Psi^+},\ket{\Phi^+}\}) &\text{~if outcome is 1} \\
				Span(\{\ket{\Psi^-},\ket{\Phi^-}\}) &\text{~if outcome is -1}.
			\end{cases}
		\end{equation}
		
		
		If we immediately then take a measurement of~$\hat{\sigma}_z\otimes \hat{\sigma}_z$ at time~$t_2 > t_1$\footnote{In the following discussions, the lower index moment always denotes earlier moments, i.e.~$t_0<t_1<t_2<t_3<\cdots$}, what can we say about the outcome probabilities \textbf{before}~$t_2$? The answer is that it depends on the initial state before~$t_1$. The unknown initial state cannot infer anything about these outcome probabilities, nor can the measurement result of~$\hat{\sigma}_x\otimes \hat{\sigma}_x$. The best we can say is that\footnote{Assume the operator~$\hat{A}$ has distinct eigenvalues~$\{a_1,a_2,\cdots,a_N\}$ then the action~``take measurement~$\hat{A}$ at time~$t$ and obtain an outcome of~$a_i$" is abbreviated as~$``\hat{A},a_i,t"$},
		\begin{equation}
			\begin{aligned}
				P(``\hat{\sigma}_z\otimes \hat{\sigma}_z,\lambda,t_2"|``\hat{\sigma}_x\otimes \hat{\sigma}_x,\lambda',t_1",I)=\text{unknown}, \\
			\end{aligned}
		\end{equation}
		where~$I$ denotes the fundamental postulates of quantum mechanics.
		
		For the sake of convenience, we could take a special choice of the initial state~$\hat{\rho}_{t=t_0}$ such that the~``unknown" is replaced with an intuitive choice,
		\begin{equation}
			\begin{aligned}
				P(``\hat{\sigma}_z\otimes \hat{\sigma}_z,\lambda,t_2"|``\hat{\sigma}_x\otimes \hat{\sigma}_x,\lambda',t_1",\hat{\rho}_{t=t_0},I)=\frac{1}{2}, \\
				P(``\hat{\sigma}_z\otimes \hat{\sigma}_z,\lambda,t_2"|\hat{\rho}_{t=t_0},I) = \frac{1}{2}.
			\end{aligned}
		\end{equation}
		
		This suggests that the measurement result of~$\hat{\sigma}_x\otimes \hat{\sigma}_x$ does nothing about the outcome probabilities of~$\hat{\sigma}_z\otimes \hat{\sigma}_z$, and if we choose an uninformative initial state, the outcome probabilities of~$\hat{\sigma}_z\otimes \hat{\sigma}_z$ before and after the measurement of~$\hat{\sigma}_x\otimes \hat{\sigma}_x$ are the same. In this situation, we may say the two measurements are independent.
		This state~$\hat{\rho}_{t=t_0}$ turns out to be a maximally entangled state for a two-body qubits system, where $\hat{\rho}_{t=t_0} = \frac{1}{4}\hat{I}\otimes\hat{I}$. Moreover, this choice is also uninformative. Under this initial state, all these global measurements will have the same outcome probabilities:
		\begin{equation}
			P(``\hat{\sigma}_{\theta,\phi}\otimes \hat{\sigma}_{\theta',\phi'},\lambda,t_1"|\hat{\rho}_{t=t_0},I) = \frac{1}{2}.
		\end{equation}
		
		We can generalize this idea to any two pairs of commuting global measurements~$\hat{\sigma}_{i_1} \otimes \hat{\sigma}_{j_1}$ and~$\hat{\sigma}_{i_2} \otimes \hat{\sigma}_{j_2}$. If we know nothing about the initial state, we could choose $\hat{\rho}_{t=t_0} = \frac{1}{4}\hat{I}\otimes\hat{I}$ to act as the prior of the initial state. This choice leads to the following relation:
		\begin{equation}
			\label{independent}
			P(``\hat{\sigma}_{i_1} \otimes \hat{\sigma}_{j_1},\lambda,t_2"|``\hat{\sigma}_{i_2} \otimes \hat{\sigma}_{j_2},\lambda',t_1",\hat{\rho}_{t=t_0},I) = \frac{1}{2}.
		\end{equation}
		
		\item $\hat{\sigma}_{i_1} \otimes \hat{\sigma}_{j_1}$ and~$\hat{\sigma}_{i_2} \otimes \hat{\sigma}_{j_2}$ are mutually unbiased
		
		Now, let's consider the example of~$\hat{\sigma}_x\otimes\hat{\sigma}_x$ and~$\hat{\sigma}_x\otimes\hat{\sigma}_z$, which do not commute, to illustrate this property. If we first take a measurement of~$\hat{\sigma}_x\otimes\hat{\sigma}_x$ at time~$t_1$ and obtain an outcome of~$+1$, the state~$\ket{\psi}_{t>t_1}$ of the system will be projected into the eigensubspace of~$\hat{\sigma}_x\otimes\hat{\sigma}_x$, denoted as $E(+1,\hat{\sigma}_x\otimes\hat{\sigma}_x)$. This eigensubspace can be expressed as a combination of Bell states:
		\begin{equation}
			\ket{\psi}_{t>t_1} = \alpha\ket{\Psi^+} + \beta\ket{\Phi^+} \quad \alpha, \beta \in \mathbb{C}, \left|\alpha\right|^2+\left|\beta\right|^2=1.
		\end{equation}
		We may also decompose~$\ket{\psi}_{t>t_1}$ as a combination of eigenstates of~$\hat{\sigma}_x\otimes\hat{\sigma}_z$:
		\begin{equation}
			\begin{aligned}
				\ket{\psi}_{t>t_1} &= \alpha\ket{\Psi^+} + \beta\ket{\Phi^+}\\
				&= \frac{\alpha}{\sqrt{2}}(\ket{0}\ket{0}+\ket{1}\ket{1}) + \frac{\beta}{\sqrt{2}}(\ket{0}\ket{1}+\ket{1}\ket{0}) \\
				&= \frac{\alpha}{2}(\ket{+}\ket{0}-\ket{-}\ket{1}) + \frac{\beta}{2}(\ket{-}\ket{1}-\ket{+}\ket{0}) +\frac{\alpha}{2}(\ket{-}\ket{0}+\ket{+}\ket{1}) + \frac{\beta}{2}(\ket{+}\ket{1}-\ket{-}\ket{0}).
			\end{aligned}
		\end{equation}
		
		This suggests after the measurement of~$\hat{\sigma}_x\otimes\hat{\sigma}_x$ with outcome~$+1$, if we take measurement~$\hat{\sigma}_x\otimes\hat{\sigma}_z$ at time~$t_2$ then the probabilities of the two outcomes of~$\hat{\sigma}_x\otimes\hat{\sigma}_z$ will be the same,
		\begin{equation}
			P(``\hat{\sigma}_x\otimes\hat{\sigma}_z,+1,t_2"|\ket{\psi}_{t>t_1},I) = P(``\hat{\sigma}_x\otimes\hat{\sigma}_z,-1,t_2"|\ket{\psi}_{t>t_1},I) = \frac{\left|\alpha\right|^2+\left|\beta\right|^2}{2} = \frac{1}{2}.
		\end{equation}
		
		The same situation happens if we obtain outcome~$-1$ by measuring~$\hat{\sigma}_x\otimes\hat{\sigma}_x$ first. More generally, if~$\hat{\sigma}_{i_1} \otimes \hat{\sigma}_{j_1}$ and~$\hat{\sigma}_{i_2} \otimes \hat{\sigma}_{j_2}$ do not commute, we will have the following relation:
		\begin{equation}
			\label{equiprobable}
			P(``\hat{\sigma}_{i_1} \otimes \hat{\sigma}_{j_1},\lambda,t_2"|``\hat{\sigma}_{i_2} \otimes \hat{\sigma}_{j_2},\lambda',t_1",h_{<t_1},I) = \frac{1}{2}~~\forall \lambda,\lambda' \in \{-1,+1\},
		\end{equation}
		where~$h_{<t_1}$ denotes all the historical measurements performed before time~$t_1$.
	\end{itemize}
	
	Both commuting and mutually unbiased global measurements yield very similar results, as revealed in the equiprobable relations of equations (\ref{independent}) and (\ref{equiprobable}). The difference is that for mutually unbiased global measurements, the equiprobable relation (\ref{equiprobable}) is valid for any prior measurement knowledge, while the two commuting operators may require a specific choice of an initial state and assume that no other measurements were performed before.
	In the following discussions, when the state of the system is not given, we tend to use the maximal mixed state as the initial state of the system. Under this initial state, both commuting and mutually unbiased measurements are non-informative; the result of one measurement cannot provide any information about the possible outcome of any subsequent measurement.

	\begin{enumerate}
		\item[] \emph{Property 4.}  In state tomography of a two-body qubit system, the outcome probability of a global measurement $\hat{\sigma}_i \otimes \hat{\sigma}_j$ can be determined by local statistics.
	\end{enumerate}

	We can always decompose the local measurements as summations of projection operators:
	\begin{align}
		\s{i}=\p{i,1}-\p{i,-1}, \s{j}=\p{j,1}-\p{j,-1}.
	\end{align}
	
	The global measurement~$\s{i}\otimes\s{j}$ can also be decomposed into summation of projectors:
	\begin{equation}
		\begin{aligned}
			\s{i}\otimes\s{j} = (\p{i,1}\otimes\p{j,1} + \p{i,-1}\otimes\p{j,-1}) - (\p{i,1}\otimes\p{j,-1} + \p{i,-1}\otimes\p{j,1}) .
		\end{aligned}
	\end{equation}

	According to L\"{u}ders' rule we may have the following relation:
	\begin{equation}
		\begin{aligned}
		P(``\s{i}\otimes\s{j},+1, t"|\hat{\rho},I) =&P(``\hat{I}\otimes\s{j},+1,t_2"|``\s{i}\otimes\hat{I},+1,t_1",\hat{\rho},I)P(``\s{i}\otimes\hat{I},+1,t_1"|\hat{\rho},I) + \\
		&P(``\hat{I}\otimes\s{j},-1,t_2"|``\s{i}\otimes\hat{I},-1,t_1",\hat{\rho},I)P(``\s{i}\otimes\hat{I},-1,t_1"|\hat{\rho},I).
		\end{aligned}    
	\end{equation}
	For any state~$\hat{\rho}$ of a two-body qubit system, the probability of obtaining~$+1$ for~$\s{i}\otimes\s{j}$ is equal to the probability of obtaining the same outcome when two local measurements are performed separately on the same system. The time order of the two local measurements is not relevant. Similarly, the probability of obtaining $-1$ for~$\s{i}\otimes\s{j}$ is equal to the probability that the two local measurements yield different outcomes.

	\subsection{Basic concepts and assumptions of question set structure}
	
	\subsubsection{Single system}
	
	We will now abstract away from this quantum description and instead regard the physical system as a black box to which we can pose one of an infinite number of different binary questions. This black box can be represented as a set $\mathcal{Q}$ of questions. Each binary outcome question $Q_{\theta, \phi} \in \mathcal{Q}$ takes the form:
	\begin{equation}
	Q_{\theta,\phi}:\text{What's the result of projective measurement }\hat{\sigma}_{\theta,\phi}\text{?}
	\end{equation}
	where
	\begin{equation}
	Q_{\theta,\phi}=0(1)\text{ if result is up (down)}.
	\end{equation}
	The interrogations of these questions are formalized as propositions. We use the following convention to express an interrogation.
	$$``Q_{\theta,\phi},q,t":\text{Conduct an interrogation of~$Q_{\theta,\phi}$ to the system at time~$t$ and obtain an outcome of~$q$} $$
	The state of the system at some time~$t$ can then be regarded as the set of the outcome probabilities of all possible propositions in~$\mathcal{Q}$ at time~$t$.
	
	In the case of binary outcomes, the two propositions, $Q_{\theta,\phi},q,t"$ and $\hat{\sigma}_{\theta,\phi},\lambda,t"$, are equivalent. We may generalize the notion of a question to correspond not just to a qubit but to an arbitrary qunit. The qunit system may also be abstracted as a black box that contains many questions, and each question~$Q$ has an outcome~$q$ in the range of~${0,1,2,\ldots, n - 1}$, which means the outcomes belong to the finite field~$\mathbb{F}_n$.
	
	Moreover, we may assume that after the interrogation of a question~$Q$, and if we keep conducting the same interrogations, the results will be the same:
	\begin{equation}
		P(``Q,q',t_2"|``Q,q,t_1",h_{<t_1},I) = \delta_{q,q'}	,
	\end{equation}
	where~$h_{<t_1}$ denotes all the historical interrogations we conducted before time~$t_1$ and~$I$ represents the basic structure of this quantum question system.
	
	We abstract the system as a set $\mathbb{Q}$, which usually contains an infinite number of questions. On one hand, we don't want to deal with an infinite degree of freedom; on the other hand, our system is taking analogies from qunits where the state has a finite number of parameters. Similarly, we may assume the question structure also has a finite number of parameters.
	
	\assumption{} For a system represented as a question set $\mathcal{Q}$, there exists a maximal subset $\mathcal{Q}_M$ containing pairwise non-informative questions, such that the outcome probability distributions of all questions in $\mathcal{Q}\setminus \mathcal{Q}_M$ are determined by the outcome probability distributions of questions in $\mathcal{Q}_M$. 
	
	\begin{figure}
		\centering
		\begin{tikzpicture}
			\filldraw[color=gray!30] (0, 0) circle(3cm);
			
			\node[draw, rectangle, minimum size=2.5cm, fill=white] (finite) at (0, 0) {};
			\node[align=center] at (finite) {};
			
			\foreach \i in {1,2,3,4,5}
			\filldraw[gray!30] (\i*72:0.8) circle (0.1) node[anchor=center] {\i};
			
			\node[right] at (45:3cm) {$\mathcal{Q}$};
			\node[] at (45:1.25cm) {$\mathcal{Q}_M$};
		\end{tikzpicture}
		\caption{\textit{Question Set Representation of Quantum System} A quantum system is abstracted as a set of questions, $\mathcal{Q}$, and it may contain an infinite number of questions. We assume that there is a finite collection of questions, $\mathcal{Q}_M$, such that the outcome probability distributions of the questions in $\mathcal{Q}_M$ determine all other outcome probabilities.}
	\end{figure}
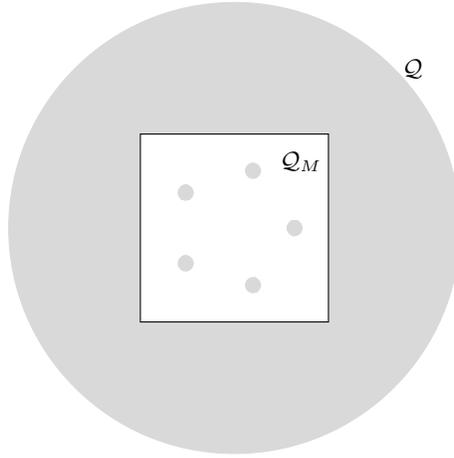
	
	Two questions, $Q_a$ and $Q_b$, are non-informative. This means that from the interrogation result of one question, we cannot obtain any information about the other question. Later, we will introduce a formal definition of information. In brief, non-informative means that if we know the interrogation result of one question and this is the only thing we know, then we cannot predict the possible result of the interrogation of the other question.
	
	In the qubit case, the subset $\mathcal{Q}_M$ is analogous to Pauli matrices. For example, $\hat{\sigma}_x$ and $\hat{\sigma}_z$ are pairwise non-informative. Once we take a measurement of $\hat{\sigma}_x$, we cannot predict the exact outcome of $\hat{\sigma}_z$. The state of the qubit could be determined via the outcome probabilities of all three operators. Although, in principle, the state of a qubit could be determined via the outcome probabilities of Stern-Gerlach projective measurements along three different axes, the three mutually perpendicular axes are more special. The Pauli matrices are mutually unbiased, making certain calculations easier. We mimic this feature in the question structure as complementary questions.
	
	
	\definition{} Two~$n$-outcome questions~$Q_a,Q_b \in \mathcal{Q}$ are said to be \textit{complementary} if
	\begin{equation}
		\begin{aligned}
			P(``Q_a,q_a,t_2"|``Q_b,q_b,t_1",h_{<t_1}, I) &= \frac{1}{n}\quad  \forall q_a,q_b\in \mathbb{F}_n, \\
			P(``Q_b,q_b,t_2"|``Q_a,q_a,t_1",h_{<t_1}, I) &= \frac{1}{n} \quad \forall q_a,q_b\in \mathbb{F}_n.
		\end{aligned}		
	\end{equation}
	In other words, no matter what interrogations have been conducted before $t_1$, the interrogation of one question will yield a uniform outcome probability distribution for the other question.
	
	Another important relation between two projective measurements is commutativity. We can think of this feature as compatible questions expressed in terms of outcome probabilities.  	
	\definition{} Two questions~$Q_a,Q_b \in \mathcal{Q}$ are said to be \textit{compatible} if
	\begin{equation}
		\begin{aligned}
			P(``Q_a, q'_a, t_3"|``Q_b,q_b,t_2",``Q_a,q_a,t_1",h_{<t_1},I) &= \delta_{q'_a,q_a}, \\
			P(``Q_b, q'_b, t_3"|``Q_a,q_a,t_2",``Q_b,q_b,t_1",h_{<t_1},I) &= \delta_{q'_b,q_b}.
		\end{aligned}
	\end{equation}
	In other words, they don't affect each other's outcomes. For any system, we first ask question $Q_a$ and obtain some outcome $q_a$. Subsequently, when we ask question $Q_b$ and obtain some outcome $q_b$, if we continue to ask question $Q_a$, we will still get outcome $q_a$, and vice versa.
	
	Compatibility is defined from an operational perspective, and while it's not exactly the same as the commutativity of operators in linear space, it is very similar. We will use compatibility as the analogy of commutativity in the following discussion.

	\subsubsection{Composite system}

	For composite systems, we can always regard them as a combination of individual subsystems. Moreover, we still treat the whole system as a set of questions, where each of them is a $d$-outcome question. It is natural to assume that the questions for a composite system contain questions from subsystems, as well as correlations between subsystems, which have a special form.
	\assumption{} Let~$\mathcal{Q}_A$ and~$\mathcal{Q}_B$ are the question sets of system~$A,B$ respectively. They form a composite system with question set~$\mathcal{Q}_{AB}$ such that
	\begin{equation}
		\mathcal{Q}_{AB}=\mathcal{Q}_A\cup \mathcal{Q}_B \cup \tilde{\mathcal{Q}}_{AB}
	\end{equation}
	$\tilde{\mathcal{Q}}_{AB}$ contains composite questions in the form $\tilde{\mathcal{Q}}_{AB} = {Q_a *_1 Q_b, Q_a *_2 Q_b, \ldots, Q_{a'} *_1 Q_{b'}, Q_{a'} *_2 Q_{b'}, \ldots}$, where $Q_a, Q_{a'}, \ldots \in \mathcal{Q}_A, Q_b, Q_{b'}, \ldots \in \mathcal{Q}_B$, and $*_1, *_2, \ldots$ are classical logical gates. Moreover, composite questions in the set $\tilde{\mathcal{Q}}_{AB}$ also have the same number of outcomes as questions in $\mathcal{Q}_A$ and $\mathcal{Q}_B$.
	
	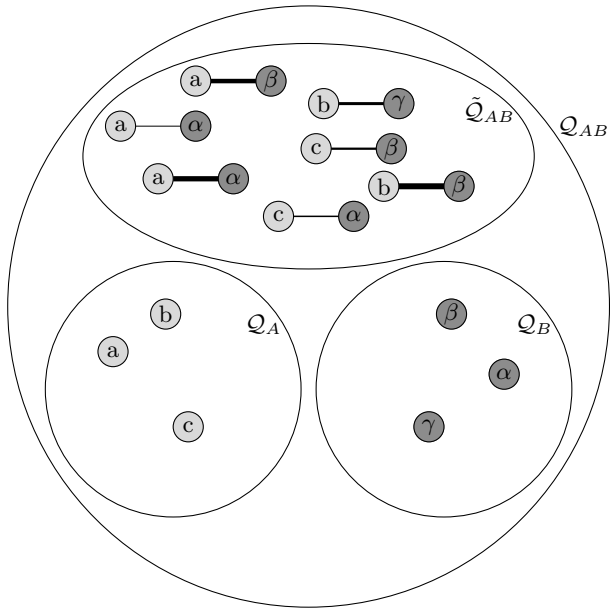
\begin{figure}[!h]
		\centering
		\begin{tikzpicture}
			\draw (0,0) ellipse (3 and 1.5);
			\filldraw[gray!30] (-2.5, 0.4) circle (0.2);
			\filldraw[gray!90] (-1.5, 0.4) circle (0.2);
			\drawCirclePair{-2.5}{0.4}{1}{0.2}{a}{$\alpha$}{line width=0.1mm}
			\filldraw[gray!30] (-2, -0.3) circle (0.2);
			\filldraw[gray!90] (-1, -0.3) circle (0.2);
			\drawCirclePair{-2}{-0.3}{1}{0.2}{a}{$\alpha$}{line width=0.7mm}
			\filldraw[gray!30] (-1.5, 1) circle (0.2);
			\filldraw[gray!90] (-0.5, 1) circle (0.2);
			\drawCirclePair{-1.5}{1}{1}{0.2}{a}{$\beta$}{line width=0.6mm}
			\filldraw[gray!30] (0.2,0.7) circle (0.2);
			\filldraw[gray!90] (1.2,0.7) circle (0.2);
			\drawCirclePair{0.2}{0.7}{1}{0.2}{b}{$\gamma$}{line width=0.4mm}
			\filldraw[gray!30] (0.1,0.1) circle (0.2);
			\filldraw[gray!90] (1.1,0.1) circle (0.2);
			\drawCirclePair{0.1}{0.1}{1}{0.2}{c}{$\beta$}{line width=0.3mm}
			\filldraw[gray!30] (-0.4,-0.8) circle (0.2);
			\filldraw[gray!90] (0.6,-0.8) circle (0.2);
			\drawCirclePair{-0.4}{-0.8}{1}{0.2}{c}{$\alpha$}{line width=0.2mm}
			\filldraw[gray!30] (1,-0.4) circle (0.2);
			\filldraw[gray!90] (2,-0.4) circle (0.2);
			\drawCirclePair{1}{-0.4}{1}{0.2}{b}{$\beta$}{line width=0.8mm}
			\node[below] at (2.4, 0.9) {$\tilde{\mathcal{Q}}_{AB}$};

			\draw (0, -2) circle (4);
			
			\draw (-1.8, -3.1) circle (1.7);
			\filldraw[gray!30] (-2.6,-2.6) circle (0.2);
			\filldraw[gray!30] (-1.9,-2.1) circle (0.2);
			\filldraw[gray!30] (-1.6,-3.6) circle (0.2);
			\drawCircle{-2.6}{-2.6}{a}{0.2}
			\drawCircle{-1.9}{-2.1}{b}{0.2}
			\drawCircle{-1.6}{-3.6}{c}{0.2}
			\node[below] at (-0.6, -2) {$\mathcal{Q}_A$};
			
			\draw (1.8, -3.1) circle (1.7);
			\filldraw[gray!90] (2.6,-2.9) circle (0.2);
			\filldraw[gray!90] (1.9,-2.1) circle (0.2);
			\filldraw[gray!90] (1.6,-3.6) circle (0.2);
			\drawCircle{2.6}{-2.9}{$\alpha$}{0.2}
			\drawCircle{1.9}{-2.1}{$\beta$}{0.2}
			\drawCircle{1.6}{-3.6}{$\gamma$}{0.2}
			\node[below] at (3.0, -2) {$\mathcal{Q}_B$};
			
			\node[right] at (3.2, 0.4) {$\mathcal{Q}_{AB}$};
		\end{tikzpicture}
		\caption{\textit{Question Set Structure in Composite System} In a composite system composed of individual systems A and B, the question set $\mathcal{Q}_{AB}$ contains both individual questions from $\mathcal{Q}_A$ and $\mathcal{Q}_B$, as well as the composite questions between these two individual systems. Each composite question is in the form of $Q_a *_i Q_b$ where $*_i$ is a logical gate, and there could be different forms of logical gates. In the figure, the different thickness of lines that connect questions represent various logical gates.}
	\end{figure}
	
	
	The correlations are also questions with outcomes in the range of $\mathbb{F}_n$. A composite question $Q_a *_i Q_b$ represents a correlation between the questions $Q_a$ and $Q_b$ in subsystems. In the case where we know the exact outcomes of $Q_a$ and $Q_b$ at the same time, denoted as $q_a$ and $q_b$ respectively, the outcome of $Q_a *_i Q_b$ is then uniquely determined by $q_a$ and $q_b$.
	
	In the qubit case, a composite question is analogous to the tensor product between two local measurements. For example, the global measurement $\hat{\sigma}_x\otimes\hat{\sigma}_z$ represents a correlation question between two local measurements $\hat{\sigma}_x\otimes \hat{I}$ and $\hat{I}\otimes\hat{\sigma}_z$. Once we know the exact outcomes of each local measurement at the same time, it means the system is in one of the eigenstates of those two local measurements. Notably, the eigenstate of the two local measurements is also the eigenstate of $\hat{\sigma}_x\otimes\hat{\sigma}_z$, making the outcome of this global measurement uniquely determined.
	
	In the qubit case, the correspondence between composite questions and global measurements may seem trivial. The non-trivial aspect lies in the existence of different forms of logical gates, meaning there could be various binary functions in the form $f:\mathbb{F}_n\times \mathbb{F}_n \rightarrow \mathbb{F}_n$. The connection between composite questions with different logical gates and operators in quantum mechanics is not obvious. We will first discuss the possible formation of these logical gates and then explore their relationship with quantum mechanics.
	
	
	
	Indeed, the degree of freedom of this composite system should be finite. The subset $\mathcal{Q}{M{AB}}$ that determines the composite system may have a structure similar to $\mathcal{Q}_{AB}$; it contains questions from subsystems as well as correlations.
	
	\assumption{} For a two-body composite system $\mathcal{Q}_{AB}$, the maximal subset $\mathcal{Q}_{M_{AB}}$ determines (the probability distribution of) outcomes of all questions in $\mathcal{Q}_{AB} \setminus \mathcal{Q}_{M_{AB}}$. $\mathcal{Q}_{M_{AB}}$ has the structure:
	\begin{equation}
	\mathcal{Q}_{M_{AB}}=\mathcal{Q}_{M_A}\cup \mathcal{Q}_{M_B} \cup \tilde{\mathcal{Q}}_{M_{AB}},
	\end{equation}
	where
	\begin{equation}
	\tilde{\mathcal{Q}}_{M_{AB}}=\{Q_a *_1 Q_b, Q_a *_2 Q_b, \cdots, Q_{a'}*_1 Q_{b'}, Q_{a'}*_2 Q_{b'},\cdots\},
	\end{equation}
	with~$Q_a,Q_{a'},\cdots \in \mathcal{Q}_{M_A}, Q_b,Q_{b'},\cdots \in \mathcal{Q}_{M_B}$. The questions in~$\tilde{\mathcal{Q}}_{M_{AB}}$ are pairwise non-informative.
	
	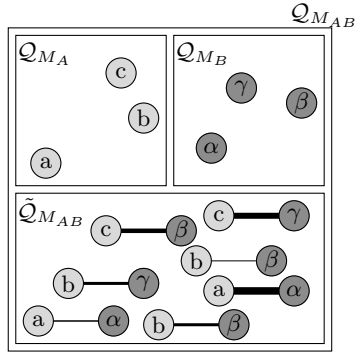
\begin{figure}[!h]
		\centering
		\begin{tikzpicture}
			\draw (-0.1, -0.1) rectangle (4.2,4.2);
			
			\draw (0,2.1) rectangle (2,4.1);
			\filldraw[gray!30] (0.4,2.4) circle (0.2);
			\filldraw[gray!30] (1.7,3.0) circle (0.2);
			\filldraw[gray!30] (1.4,3.6) circle (0.2);
			\drawCircle{0.4}{2.4}{a}{0.2}
			\drawCircle{1.7}{3.0}{b}{0.2}
			\drawCircle{1.4}{3.6}{c}{0.2}
			
			\draw (2.1,2.1) rectangle (4.1,4.1);
			\filldraw[gray!90] (2.6,2.6) circle (0.2);
			\filldraw[gray!90] (3.8,3.2) circle (0.2);
			\filldraw[gray!90] (3,3.4) circle (0.2);
			\drawCircle{2.6}{2.6}{$\alpha$}{0.2}
			\drawCircle{3.8}{3.2}{$\beta$}{0.2}
			\drawCircle{3.0}{3.4}{$\gamma$}{0.2}
			
			\draw (0,0) rectangle (4.1,2);
			\filldraw[gray!30] (0.3,0.3) circle (0.2);
			\filldraw[gray!90] (1.3,0.3) circle (0.2);
			\drawCirclePair{0.3}{0.3}{1}{0.2}{a}{$\alpha$}{line width=0.2mm}
			\filldraw[gray!30] (0.7,0.8) circle (0.2);
			\filldraw[gray!90] (1.7,0.8) circle (0.2);
			\drawCirclePair{0.7}{0.8}{1}{0.2}{b}{$\gamma$}{line width=0.4mm}
			\filldraw[gray!30] (1.2,1.5) circle (0.2);
			\filldraw[gray!90] (2.2,1.5) circle (0.2);
			\drawCirclePair{1.2}{1.5}{1}{0.2}{c}{$\beta$}{line width=0.6mm}
			\filldraw[gray!30] (2.7,1.7) circle (0.2);
			\filldraw[gray!90] (3.7,1.7) circle (0.2);
			\drawCirclePair{2.7}{1.7}{1}{0.2}{c}{$\gamma$}{line width=0.8mm}
			\filldraw[gray!30] (2.4,1.1) circle (0.2);
			\filldraw[gray!90] (3.4,1.1) circle (0.2);
			\drawCirclePair{2.4}{1.1}{1}{0.2}{b}{$\beta$}{line width=0.0mm}
			\filldraw[gray!30] (2.7,0.7) circle (0.2);
			\filldraw[gray!90] (3.7,0.7) circle (0.2);
			\drawCirclePair{2.7}{0.7}{1}{0.2}{a}{$\alpha$}{line width=1.0mm}
			\filldraw[gray!30] (1.9,0.25) circle (0.2);
			\filldraw[gray!90] (2.9,0.25) circle (0.2);
			\drawCirclePair{1.9}{0.25}{1}{0.2}{b}{$\beta$}{line width=0.4mm}
			
			\node[right] at (-0.1, 3.85) {$\mathcal{Q}_{M_A}$};
			\node[right] at (2.0, 3.85) {$\mathcal{Q}_{M_B}$};
			\node[right] at (-0.1, 1.75) {$\tilde{\mathcal{Q}}_{M_{AB}}$};
			\node[above] at (4.1, 4.1) {${\mathcal{Q}}_{M_{AB}}$};
		\end{tikzpicture}
		\caption{\textit{$\mathcal{Q}_M$ of Composite System} We assume that the subset $\mathcal{Q}{M}$ of a composite system, which is formed by combining individual systems A and B, exhibits a structure akin to the question set $\mathcal{Q}{AB}$. $\mathcal{Q}{M}$ includes questions from both $\mathcal{Q}{M_A}$ and $\mathcal{Q}_{M_B}$, along with composite questions formed using various types of logical gates.}
	\end{figure}
	
	So far, we know little about the properties of these correlation questions and logical gates. We will first investigate the construction of these logical gates using the restriction that questions in $\mathcal{Q}_{M_{AB}}$ are pairwise non-informative. After that, we may propose information about questions and use assumptions on information to derive the detailed structure of the subset $\mathcal{Q}_{M}$ for both single systems and composite systems.

	\subsection{Restrictions of logical gates and generalizations in higher dimension}
	
	The key aspect to investigate in the structure of logical gates is the assumption that the questions in $\mathcal{Q}_{M_{AB}}$ are pairwise non-informative. Now, consider four questions in a composite system: $Q_a, Q_b, Q_a *i Q_b, Q_a *j Q_b \in \mathcal{Q}_{M_{AB}}$. They are pairwise non-informative. This implies there are two restrictions on the choices of logic gates:
	\begin{enumerate}
		\item[] \emph{Restriction 1.}  $Q_a *_i Q_b$ is non-informative to both~$Q_a,Q_b$ respectively;
		\item[] \emph{Restriction 2.} $Q_a *_i Q_b$ is non-informative to~$Q_a *_j Q_b$ if~$*_i$ and~$*_j$ are different logical gates.
	\end{enumerate}
	
	In the case of binary outcome systems, the choice of $*_i$ is unique (see Table~\ref{tbl:binary-gates}). There are a total of 16 binary logical gates, while only XNOR or XOR satisfy the first restriction. Although XNOR and XOR are different, they don't satisfy the second restriction. Notice that if we know the value of $Q_a$ XNOR $Q_b$, we can immediately infer the value of $Q_a$ XOR $Q_b$, and vice versa. Therefore, they are equivalent up to a negation operation.
	
	\begin{table}[!h]
		\centering
		\begin{tabular}{|c|c|c|}
			\hline
			$Q_a$ & $Q_b$ & $Q_a$ AND $Q_b$ \\ \hline
			0     & 0     & 0               \\ \hline
			0     & 1     & 0               \\ \hline
			1     & 0     & 0               \\ \hline
			1     & 1     & 1               \\ \hline
		\end{tabular}
		\begin{tabular}{|c|c|c|}
			\hline
			$Q_a$ & $Q_b$ & $Q_a$ OR $Q_b$ \\ \hline
			0     & 0     & 0               \\ \hline
			0     & 1     & 1               \\ \hline
			1     & 0     & 1               \\ \hline
			1     & 1     & 1               \\ \hline
		\end{tabular}
		\begin{tabular}{|c|c|c|}
			\hline
			$Q_a$ & $Q_b$ & $Q_a$ XOR $Q_b$ \\ \hline
			0     & 0     & 0               \\ \hline
			0     & 1     & 1               \\ \hline
			1     & 0     & 1               \\ \hline
			1     & 1     & 0               \\ \hline
		\end{tabular}
		\begin{tabular}{|c|c|c|}
			\hline
			$Q_a$ & $Q_b$ & $Q_a$ XNOR $Q_b$ \\ \hline
			0     & 0     & 1               \\ \hline
			0     & 1     & 0               \\ \hline
			1     & 0     & 0               \\ \hline
			1     & 1     & 1               \\ \hline
		\end{tabular}
		\caption{\label{tbl:binary-gates}\textit{Binary Logical Gates Exmaples} The AND gate does not satisfy the first restriction, since if we know the result of~$Q_a$ AND $Q_b$ is~$1$, then we immediately know both~$Q_a$ and~$Q_b$ must have outcome~$1$ which is not non-informative. The OR gate meets from a similar problem.  Of the 16 possible two-input one-output logic gates, only XOR and XNOR satisfy the first restriction.} 
	\end{table}
	
	
	
	The logical gates in the~$n$-ary case are more complex than binary logic gates. In binary case there is only one allowable logical gate which is XOR (or XNOR up to a negation). But, in the~$n$-ary case, there could be more than one allowable logical gate.  Table~\ref{tbl:ternary-gates} shows an example of logical gates in the ternary case.
	
	
	\begin{table}[!h]
		\centering
		\begin{tabular}{|c|c|c|}
			\hline
			$Q_a$ & $Q_b$ & $Q_a$$\times$$Q_b$ \\ \hline
			0 & 0 & 0       \\ 
			2 & 1 & 2       \\ 
			1 & 2 & 2		\\ \hline
			2 & 0 & 0       \\ 
			1 & 1 & 1       \\ 
			0 & 2 & 0		\\ \hline
			1 & 0 & 0		\\ 
			0 & 1 & 0		\\ 
			2 & 2 & 1		\\ \hline
		\end{tabular} \quad
		\begin{tabular}{|c|c|c|}
			\hline
			$Q_a$ & $Q_b$ & $Q_a+Q_b$ \\ \hline
			0 & 0 & 0       \\ 
			2 & 1 & 0       \\ 
			1 & 2 & 0		\\ \hline
			1 & 0 & 1		\\ 
			0 & 1 & 1		\\ 
			2 & 2 & 1		\\ \hline
			2 & 0 & 2       \\ 
			1 & 1 & 2       \\ 
			0 & 2 & 2		\\ \hline
		\end{tabular} \quad
		\begin{tabular}{|c|c|c|}
			\hline
			$Q_a$ & $Q_b$ & $Q_a-Q_b$ \\ \hline
			0 & 0 & 0       \\ 
			1 & 1 & 0       \\ 
			2 & 2 & 0		\\ \hline
			1 & 0 & 1       \\ 
			2 & 1 & 1       \\ 
			0 & 2 & 1		\\ \hline
			2 & 0 & 2		\\ 
			0 & 1 & 2		\\ 
			1 & 2 & 2		\\ \hline
		\end{tabular}
		\caption{\label{tbl:ternary-gates}\textit{Ternary Logical Gates Exmaples} Ternary multiplication does not satisfy restriction 1 since if we know the value of~$Q_a\times Q_b$ is non-zero we immediately know the values of both~$Q_a$ and~$Q_b$ cannot be zero. While ternary addition and ternary subtraction satisfy both two restrictions. }
	\end{table}
	
	In ternary case there are at least two different logical gates satisfy restriction 1. Before discussing different logical gates, we first exclude equivalent logical gates.
	%
	\begin{table}[!h]
		\begin{center}
			\begin{tabular}{|l|l|c|}
				\hline
				$Q_a$ & $Q_b$ & $Q_a+Q_b$ \\ \hline
				0 & 0 & \textcolor{red}{0}       \\ 
				2 & 1 & \textcolor{red}{0}       \\ 
				1 & 2 & \textcolor{red}{0}		\\ \hline
				1 & 0 & \textcolor{blue}{1}		\\ 
				0 & 1 & \textcolor{blue}{1}		\\ 
				2 & 2 & \textcolor{blue}{1}		\\ \hline
				2 & 0 & \textcolor{green}{2}       \\ 
				1 & 1 & \textcolor{green}{2}       \\ 
				0 & 2 & \textcolor{green}{2}		\\ \hline
			\end{tabular} \quad $\xrightarrow{0 \rightarrow 2, 1 \rightarrow 0, 2 \rightarrow 1}$
			\begin{tabular}{|l|l|c|}
				\hline
				$Q_a$ & $Q_b$ & $Q_a*_+Q_b$ \\ \hline
				0 & 0 & \textcolor{green}{2}       \\ 
				2 & 1 & \textcolor{green}{2}       \\ 
				1 & 2 & \textcolor{green}{2}		\\ \hline
				2 & 0 & \textcolor{red}{0}       \\ 
				1 & 1 & \textcolor{red}{0}       \\ 
				0 & 2 & \textcolor{red}{0}		\\ \hline
				1 & 0 & \textcolor{blue}{1}		\\ 
				0 & 1 & \textcolor{blue}{1}		\\ 
				2 & 2 & \textcolor{blue}{1}		\\ \hline
			\end{tabular}
			\caption{\label{tbl:ternary-addition-variant}\textit{Variant of Ternary Addition} By knowing the value of~$Q_a+Q_b$ we can immediately know the value of~$Q_a*_+Q_b$, and vice versa. These two logical gates are equivalent up to a permutation~$(021)$.}
		\end{center}
	\end{table}
	In the example shown in Table~\ref{tbl:ternary-addition-variant}, if we know the value of $Q_a+Q_b$, then we know the value of $Q_a *_+ Q_b$, and vice versa. Just as in the binary case, where XNOR and XOR are equivalent up to a negation operation, the operations $Q_a+Q_b$ and $Q_a *_+ Q_b$ are also equivalent up to a permutation in the symmetric group $S_3$. This can be generalized to $n$-ary logical gates: if an $n$-ary logical gate satisfies restriction 1, then it is equivalent to $n!-1$ variations, with each variation corresponding to a non-identity element in $S_n$.
	
	Now, we can investigate different logical gates, and two problems need to be solved. First, what is the maximal number of different logical gates in the $n$-ary case? And how can we construct these different logical gates?
	
	By excluding equivalent variations, the truth table of each logical gate can be written in the same pattern. As Table~\ref{tbl:example_compl_oper} shows, the truth table of a logical gate is divided into blocks, and in each block, the second column ranges orderly in ${0,1,\ldots,n-1}$, and the first column remains constant.
	
	\begin{table}[!h]
		\centering
		\begin{tabular}{|c|c|c|}
			\hline
			$Q_a$ & $Q_b$ & $Q_a+Q_b$ \\ \hline
			0 & 0 & 0       \\ 
			0 & 1 & 1       \\ 
			0 & 2 & 2		\\ \hline
			1 & 0 & 1		\\ 
			1 & 1 & 2		\\ 
			1 & 2 & 0		\\ \hline
			2 & 0 & 2       \\ 
			2 & 1 & 0       \\ 
			2 & 2 & 1		\\ \hline
		\end{tabular} \qquad
		\begin{tabular}{|c|c|c|}
			\hline
			$Q_a$ & $Q_b$ & $Q_a-Q_b$ \\ \hline
			0 & 0 & 0       \\ 
			0 & 1 & 2       \\ 
			0 & 2 & 1		\\ \hline
			1 & 0 & 1       \\ 
			1 & 1 & 0       \\ 
			1 & 2 & 2		\\ \hline
			2 & 0 & 2		\\ 
			2 & 1 & 1		\\ 
			2 & 2 & 0		\\ \hline
		\end{tabular} \qquad
		\begin{tabular}{|c|c|c|}
			\hline
			$Q_a$ & $Q_b$ & $Q_a*_1Q_b$ \\ \hline
			0 & 0 & 0   \\ 
			0 & 1 & 1   \\ 
			0 & 2 & 2   \\ 
			0 & 3 & 3   \\ 
			0 & 4 & 4   \\ \hline
			1 & 0 & 1   \\ 
			1 & 1 & 2   \\ 
			1 & 2 & 3   \\ 
			1 & 3 & 4   \\ 
			1 & 4 & 0   \\ \hline
			2 & 0 & 2   \\ 
			2 & 1 & 3   \\ 
			2 & 2 & 4   \\ 
			2 & 3 & 0   \\ 
			2 & 4 & 1   \\ \hline
			3 & 0 & 3   \\ 
			3 & 1 & 4   \\ 
			3 & 2 & 0   \\ 
			3 & 3 & 1   \\ 
			3 & 4 & 2   \\ \hline
			4 & 0 & 4   \\ 
			4 & 1 & 0   \\ 
			4 & 2 & 1   \\ 
			4 & 3 & 2   \\ 
			4 & 4 & 3   \\ \hline
		\end{tabular} \qquad
		\begin{tabular}{|c|c|c|}
			\hline
			$Q_a$ & $Q_b$ & $Q_a*_2Q_b$ \\ \hline
			0 & 0 & 0   \\ 
			0 & 1 & 2   \\ 
			0 & 2 & 4   \\ 
			0 & 3 & 1   \\ 
			0 & 4 & 3   \\ \hline
			1 & 0 & 1   \\ 
			1 & 1 & 3   \\ 
			1 & 2 & 0   \\ 
			1 & 3 & 2   \\ 
			1 & 4 & 4   \\ \hline
			2 & 0 & 2   \\ 
			2 & 1 & 4   \\ 
			2 & 2 & 1   \\ 
			2 & 3 & 3   \\ 
			2 & 4 & 0   \\ \hline
			3 & 0 & 3   \\ 
			3 & 1 & 0   \\ 
			3 & 2 & 2   \\ 
			3 & 3 & 4   \\ 
			3 & 4 & 1   \\ \hline
			4 & 0 & 4   \\ 
			4 & 1 & 1   \\ 
			4 & 2 & 3   \\ 
			4 & 3 & 0   \\ 
			4 & 4 & 2   \\ \hline
		\end{tabular}
		\caption{\label{tbl:example_compl_oper} \textit{Example of Logical gates by Fixing Two Columns} We rearrange the first two columns of a logical gate in the given order. For a $n$-ary logical gates, we divide the truth table into $n$ different blocks. In the first column, each block contains only one number while in the second column each block contains numbers from $0$ to $n-1$.}
	\end{table}
	
	The benefit of this pattern is that we can combine different tables into one larger table. By doing so, the combined table becomes an orthogonal array, a concept well-investigated in mathematics. Table~\ref{tbl:combined-table-ternary} provides an example of a combined table in the ternary case.
	
	\begin{table}[!h]
		\centering
		\begin{tabular}{|c|c|c|}
			\hline
			$Q_a$ & $Q_b$ & $Q_a+Q_b$ \\ \hline
			0 & 0 & 0       \\ 
			0 & 1 & 1       \\ 
			0 & 2 & 2		\\ \hline
			1 & 0 & 1		\\ 
			1 & 1 & 2		\\ 
			1 & 2 & 0		\\ \hline
			2 & 0 & 2       \\ 
			2 & 1 & 0       \\ 
			2 & 2 & 1		\\ \hline
		\end{tabular} $\cup$
		\begin{tabular}{|c|c|c|}
			\hline
			$Q_a$ & $Q_b$ & $Q_a-Q_b$ \\ \hline
			0 & 0 & 0       \\ 
			0 & 1 & 2       \\ 
			0 & 2 & 1		\\ \hline
			1 & 0 & 1       \\ 
			1 & 1 & 0       \\ 
			1 & 2 & 2		\\ \hline
			2 & 0 & 2		\\ 
			2 & 1 & 1		\\ 
			2 & 2 & 0		\\ \hline
		\end{tabular} $\rightarrow$
		\begin{tabular}{|c|c|c|c|}
			\hline
			$Q_a$ & $Q_b$ & $Q_a + Q_b$ & $Q_a - Q_b$ \\ \hline
			0  & 0  & 0 & 0      \\ 
			0  & 1  & 1 & 2      \\ 
			0  & 2  & 2 & 1      \\ \hline
			1  & 0  & 1 & 1      \\ 
			1  & 1  & 2 & 0      \\ 
			1  & 2  & 0 & 2      \\ \hline
			2  & 0  & 2 & 2      \\ 
			2  & 1  & 0 & 1      \\ 
			2  & 2  & 1 & 0      \\ \hline
		\end{tabular}
		\caption{\label{tbl:combined-table-ternary}\textit{Logical Gate and Orthogonal Array} The truth tables of ternary addition and ternary subtraction are both orthogonal arrays of 9 rows , 3 columns, level 3 and strength 2. Level 3 means there are 3 different elements. Strength 2 means it is a table of 9 rows and 3 columns and for every selection of 2 columns, all ordered 2-tuples of the elements appear exactly $\frac{row}{level^{strength}}$ times. These two tables can be combined into a larger orthogonal array with 4 columns.}
	\end{table}
	
	The combined table of ternary addition and subtraction is an \textit{orthogonal array}~\cite{Rao2009, OA}. An orthogonal array, denoted as $OA(N,k,s,t)$, is an array with $N$ rows and $k$ columns, where there are $s$ different elements, and its strength is $t$. This means that every $N\times t$ subarray contains each $t$-tuple exactly $\lambda$ times as a row, with $\lambda=N/s^t$. In this case, the orthogonal array is of size $3^2\times 4$, with level 3 and strength 2.
	
	As all possible logical gates can be combined into a single orthogonal array, the question of the maximal number of different logical gates can be rephrased as follows: What is the maximal number of columns in an orthogonal array with size $n^2$, level $n$, and strength 2? This problem is well-investigated when $n$ is a power of a prime number but extremely difficult in other cases.
	
	\fact{} The maximal number of columns for an orthogonal array with $n^2$ rows, level $n$, and strength 2 is $n+1$ if $n$ is a prime power~\cite[p.~38]{OA}.
	
	This fact implies that when $n$ is a prime power, there will be $n-1$ different logical gates. The next problem is the construction of these $n-1$ different logical gates. We have found that, if $n$ is a prime number, the $n-1$ logical gates, denoted as ${_1,2,\cdots,*{n-1}}$, can be represented in this way:
	\begin{equation}
	Q_a*_iQ_b:=Q_a+i\times Q_b \quad (\text{mod } n)\quad \forall i \in \{1,2,\cdots,n-1\}.
	\end{equation}
	The example of the quinary case is shown in Table~\ref{tbl:quinary-gates}.
	
	\begin{table}[!h]
		\centering
		\begin{tabular}{|c|c|c|}
			\hline
			$Q_a$ & $Q_b$ & $Q_a+Q_b$ \\ \hline
			\textcolor{red}{0} & \textcolor{red}{0} & \textcolor{red}{0}   \\ 
			0 & 1 & 1   \\ 
			0 & 2 & 2   \\ 
			0 & 3 & 3   \\ 
			0 & 4 & 4   \\ \hline
			1 & 0 & 1   \\ 
			1 & 1 & 2   \\ 
			1 & 2 & 3   \\ 
			1 & 3 & 4   \\ 
			\textcolor{orange}{1} & \textcolor{orange}{4} & \textcolor{orange}{0}   \\ \hline
			2 & 0 & 2   \\ 
			2 & 1 & 3   \\ 
			2 & 2 & 4   \\ 
			\textcolor{green}{2} & \textcolor{green}{3} & \textcolor{green}{0}   \\ 
			2 & 4 & 1   \\ \hline
			3 & 0 & 3   \\ 
			3 & 1 & 4   \\ 
			\textcolor{blue}{3} & \textcolor{blue}{2} & \textcolor{blue}{0}   \\ 
			3 & 3 & 1   \\ 
			3 & 4 & 2   \\ \hline
			4 & 0 & 4   \\ 
			\textcolor{purple}{4} & \textcolor{purple}{1} & \textcolor{purple}{0}   \\ 
			4 & 2 & 1   \\ 
			4 & 3 & 2   \\ 
			4 & 4 & 3   \\ \hline
		\end{tabular} \quad
		\begin{tabular}{|c|c|c|}
			\hline
			$Q_a$ & $Q_b$ & $Q_a+2\times Q_b$ \\ \hline
			\textcolor{red}{0} & \textcolor{red}{0} & \textcolor{red}{0}            \\ 
			0 & 1 & 2            \\ 
			0 & 2 & 4            \\ 
			0 & 3 & 1            \\ 
			0 & 4 & 3            \\ \hline
			1 & 0 & 1            \\ 
			1 & 1 & 3            \\ 
			1 & 2 & 0            \\ 
			1 & 3 & 2            \\ 
			\textcolor{orange}{1} & \textcolor{orange}{4} & \textcolor{orange}{4}            \\ \hline
			2 & 0 & 2            \\ 
			2 & 1 & 4            \\ 
			2 & 2 & 1            \\ 
			\textcolor{green}{2} & \textcolor{green}{3} & \textcolor{green}{3}            \\ 
			2 & 4 & 0            \\ \hline
			3 & 0 & 3            \\ 
			3 & 1 & 0            \\ 
			\textcolor{blue}{3} & \textcolor{blue}{2} & \textcolor{blue}{2}            \\ 
			3 & 3 & 4            \\ 
			3 & 4 & 1            \\ \hline
			4 & 0 & 4            \\ 
			\textcolor{purple}{4} & \textcolor{purple}{1} & \textcolor{purple}{1}            \\ 
			4 & 2 & 3            \\ 
			4 & 3 & 0            \\ 
			4 & 4 & 2            \\ \hline
		\end{tabular}\quad
		\begin{tabular}{|c|c|c|}
			\hline
			$Q_a$ & $Q_b$ & $Q_a+3\times Q_b$ \\ \hline
			\textcolor{red}{0} & \textcolor{red}{0} & \textcolor{red}{0}   \\ 
			0 & 1 & 3   \\ 
			0 & 2 & 1   \\ 
			0 & 3 & 4   \\ 
			0 & 4 & 2   \\ \hline
			1 & 0 & 1   \\ 
			1 & 1 & 4   \\ 
			1 & 2 & 2   \\ 
			1 & 3 & 0   \\ 
			\textcolor{orange}{1} & \textcolor{orange}{4} & \textcolor{orange}{3}   \\ \hline
			2 & 0 & 2   \\ 
			2 & 1 & 0   \\ 
			2 & 2 & 3   \\ 
			\textcolor{green}{2} & \textcolor{green}{3} & \textcolor{green}{1}   \\ 
			2 & 4 & 4   \\ \hline
			3 & 0 & 3   \\ 
			3 & 1 & 1   \\ 
			\textcolor{blue}{3} & \textcolor{blue}{2} & \textcolor{blue}{4}   \\ 
			3 & 3 & 2   \\ 
			3 & 4 & 0   \\ \hline
			4 & 0 & 4   \\ 
			\textcolor{purple}{4} & \textcolor{purple}{1} & \textcolor{purple}{2}   \\ 
			4 & 2 & 0   \\ 
			4 & 3 & 3   \\ 
			4 & 4 & 1   \\ \hline
		\end{tabular} \quad
		\begin{tabular}{|c|c|c|}
			\hline
			$Q_a$ & $Q_b$ & $Q_a+4\times Q_b$ \\ \hline
			\textcolor{red}{0} & \textcolor{red}{0} & \textcolor{red}{0}   \\ 
			0 & 1 & 4   \\ 
			0 & 2 & 3   \\ 
			0 & 3 & 2   \\ 
			0 & 4 & 1   \\ \hline
			1 & 0 & 1   \\ 
			1 & 1 & 0   \\ 
			1 & 2 & 4   \\ 
			1 & 3 & 3   \\ 
			\textcolor{orange}{1} & \textcolor{orange}{4} & \textcolor{orange}{2}   \\ \hline
			2 & 0 & 2   \\ 
			2 & 1 & 1   \\ 
			2 & 2 & 0   \\ 
			\textcolor{green}{2} & \textcolor{green}{3} & \textcolor{green}{4}   \\ 
			2 & 4 & 3   \\ \hline
			3 & 0 & 3   \\ 
			3 & 1 & 2   \\ 
			\textcolor{blue}{3} & \textcolor{blue}{2} & \textcolor{blue}{1}   \\ 
			3 & 3 & 0   \\ 
			3 & 4 & 4   \\ \hline
			4 & 0 & 4   \\ 
			\textcolor{purple}{4} & \textcolor{purple}{1} & \textcolor{purple}{3}   \\ 
			4 & 2 & 2   \\ 
			4 & 3 & 1   \\ 
			4 & 4 & 0   \\ \hline
		\end{tabular}
		\caption{\label{tbl:quinary-gates} \textit{Four Different Quinary Logical Gates} The colored numbers indicates how the four logical gates are non-informative to each other. If the value of~$Q_a + Q_b$ is~$0$, there will be five different combinations of values of~$Q_a$ and~$Q_b$, and each combination yields a different value in other gate.}
	\end{table}
	
	Unfortunately, this construction fails when $n$ is a prime power. Up to now, we have not found any elegant representations of logical gates when $n$ is a prime power. Therefore, we shall henceforth focus on the case where all questions have prime number outcomes.
	
	Once we obtain a relatively clear form of logical gates in a $p$-ary system, we immediately have the following consequence:
	
	\cor{} In a two-body composite system $\mathcal{Q}_{AB}$, $Q_a, Q_b, Q_a * Q_b$ are mutually compatible, where $Q_a\in \mathcal{Q}_A, Q_b \in \mathcal{Q}_B$, and $*$ is any allowable logical gate. If the exact values of two of them are known, then the value of the remaining question will be ensured due to the specific form of logical gate $*$.
	
	This result is an analogy of the commutativity between $\hat{\sigma}_i\otimes\hat{I},\hat{I}\otimes\hat{\sigma}_j, \hat{\sigma}_i\otimes\hat{\sigma}_j$. If we take measurements of any two of the three operators, the state of the system will be ensured, and the possible outcome of the unmeasured operator can also be ensured.
	
	In the binary case, since there is only one allowable logical gate, the relation between composite questions and composite operators seems natural when replacing $*$ with $\otimes$. In higher-order cases when we have more allowable logical gates, this correspondence may not be very clear. Later, we will introduce a general correspondence between composite questions and composite operators.
	
	\subsection{Information of questions and consequences}
	
	So far, we have obtained a nice property and expression of logical gates, specifically in prime number dimensional systems. However, this couldn't yield more information about the internal structure of $\mathcal{Q}_M$. Suppose we assume that $\mathcal{Q}_M$ contains a finite number of questions, but what is that number? In the following discussions, we will introduce a new concept: information of questions, to help construct the structure of quantum questions under informational postulates.
	
	Like many existing information measures that are built upon probability distributions, we tend to define the information of a question based on its outcome probability. Given the background knowledge of the system, the historical interrogations $h_{<t}$ we have conducted on the system before time $t$, the outcome probability of a question $Q$ at this moment $t$ is denoted as $P(``Q,q,t"|h_{<t},I)$. For an $n$-outcome question, there will be $n$ outcome probabilities.
	
	The information of a question $Q$ given the background knowledge $h_{<t}$ is a function of these outcome probabilities:
	\begin{equation}
		I(Q|h_{<t}) = H(P(``Q,0,t"|h_{<t},I), P(``Q,1,t"|h_{<t},I), \cdots, P(``Q,n-1,t"|h_{<t},I)).
	\end{equation}

	$H$ is a function of probability distributions and as a convention we set the following restrictions of information~$I$ and function~$H$:
	\begin{enumerate}
		\item $0\le I(Q|h_{<t})\le 1$;
		\item $H(\vec{p}) = 0$ if and only if~$\vec{p} = (\frac{1}{n},\frac{1}{n},\cdots,\frac{1}{n})$;
		\item $H(\vec{p}) = 1$ if and only if~$\vec{p}$ is a~$n$-tuple contains~$n-1$ zeros and~$1$ one.
	\end{enumerate}

	Instead of offering a formal measure of information, our approach focuses on considering two extreme cases concerning the information of questions.
	To find a detailed expression of this information, we would need more informational assumptions regarding the constraints of this measure. However, at the current stage, we cannot find more intuitive informational postulates.
	In fact, in the following calculations, the information of each question is either 0 or 1. The two extreme cases are already sufficient for us to demonstrate the subtle structure of quantum questions.
	
	Now, we can define non-informative questions in terms of information. Two questions, $Q_a$ and $Q_b$, are said to be pairwise non-informative if from the interrogation of one of them alone, we cannot obtain any information about the other question,
	\begin{equation}
		I(Q_a|``Q_b,q,t") = 0, \quad I(Q_b|``Q_a,q,t") = 0.
	\end{equation}

	This definition lead to two subsequent consequences on complementary and compatible questions.
	
	\cor{} If we perform interrogations on two complementary questions, the latter interrogation will erase the information about the question interrogated earlier.
	\begin{proof}
		Assume two questions~$Q_a$ and~$Q_b$ are pairwise complementary. If we take interrogation of question~$Q_a$ at time~$t_1$ with outcome~$q_a$, then after the interrogation we obtain 1 unit information of~$Q_a$ and 0 unit information of~$Q_b$:
		\begin{equation}
			\begin{aligned}
				P(``Q_a, q'_a, t_2"|``Q_a, q_a, t_1",I) &= \delta_{q'_a, q_a}, &&\quad I(Q_a|``Q_a, q_a, t_1",I) = 1;\\
				P(``Q_b, q_b, t_2"|``Q_a, q_a, t_1",I) &= \frac{1}{n}, && \quad I(Q_b|``Q_a, q_a, t_1",I) = 0.
			\end{aligned}
		\end{equation}
		Then we take interrogation of question~$Q_b$ at time~$t_2$ with outcome~$q_b$. After~$t_2$ we gain 1 unit information about question~$Q_b$ and 0 unit information about~$Q_a$:
		\begin{equation}
			\begin{aligned}
				&P(``Q_a, q'_a, t_2"|``Q_b,q_b,t_2",``Q_a, q_a, t_1",I) = \frac{1}{n},\quad &&I(Q_a|``Q_b,q_b,t_2",``Q_a, q_a, t_1",I) = 0; \\
				&P(``Q_b, q'_b, t_2"|``Q_b,q_b,t_2",``Q_a, q_a, t_1",I) = \delta_{q'_b, q_b},\quad &&I(Q_b|``Q_b,q_b,t_2",``Q_a, q_a, t_1",I) = 1.
			\end{aligned}
		\end{equation}
		
		This shows the interrogation~$``Q_b,q_b,t_2"$ erases the information we gain about~$Q_a$ at time~$t_1$. Similar results will be yielded if we change the order of interrogations on~$Q_a$ and~$Q_b$.
	\end{proof}

	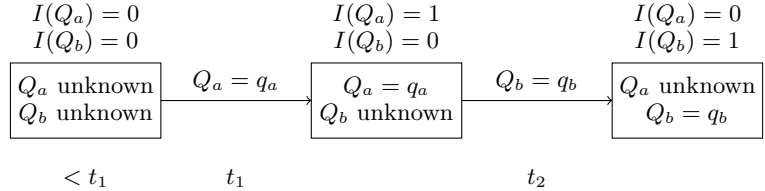
\begin{figure}[!h]
		\centering
		\begin{tikzpicture}
			
			\draw (1,2) rectangle (3,3) node[pos=.5, align=center] {$Q_a$ unknown\\$Q_b$ unknown};
			\node[above, align=center] at (2,3) {$I(Q_a)=0$\\$I(Q_b)=0$};
			
			\draw (5,2) rectangle (7,3) node[pos=.5,align=center] {$Q_a = q_a$\\$Q_b$ unknown};
			\node[above, align=center] at (6,3) {$I(Q_a)=1$\\$I(Q_b)=0$};
			
			\draw (9,2) rectangle (11,3) node[pos=.5,align=center] {$Q_a$ unknown\\$Q_b=q_b$};
			\node[above, align=center] at (10,3) {$I(Q_a)=0$\\$I(Q_b)=1$};
			
			\draw[->] (3,2.5) -- node[above] {$Q_a=q_a$} (5,2.5);
			\draw[->] (7,2.5) -- node[above] {$Q_b=q_b$} (9,2.5);
			
			\node[below, align=center] at (2,1.7) {$<t_1$};
			\node[below, align=center] at (4,1.7) {$t_1$};
			\node[below, align=center] at (8,1.7) {$t_2$};
			
		\end{tikzpicture}
		\caption{\textit{Mutually Complementary Questions in Interrogation} $Q_a$ and $Q_b$ are mutually complementary. Initially, we have no knowledge about the state and two questions. At time $t_1$, we conduct an interrogation of $Q_a$ with an outcome $q_a$. After this interrogation, we gain 1 unit of information about $Q_a$ and cannot obtain any information about $Q_b$. At time $t_2$, we conduct another interrogation of $Q_b$ with outcome $q_b$, and this interrogation may erase the information of $Q_a$.}
	\end{figure}
	
	\cor{} If we perform interrogations on two compatible questions, the later interrogation will retain the information about the question interrogated earlier.
	\begin{proof}
		For two compatible questions~$Q_a$ and~$Q_b$, from the definition we may have the following relations:
		\begin{equation}
			\begin{aligned}
				P(``Q_a, q'_a, t_3"|``Q_b,q_b,t_2",``Q_a,q_a,t_1",h_{<t_1},I) &= \delta_{q'_a,q_a}, \\
				P(``Q_b, q'_b, t_3"|``Q_a,q_a,t_2",``Q_b,q_b,t_1",h_{<t_1},I) &= \delta_{q'_b,q_b}.
			\end{aligned}
		\end{equation}
		This suggests that~$I(Q_a|``Q_b,q_b,t_2",``Q_a,q_a,t_1",h_{<t_1}) = I(Q_b|``Q_a,q_a,t_2",``Q_b,q_b,t_1",h_{<t_1}) = 1$. If we have conducted an interrogation of a question, the latter interrogation of another compatible question will retain the information obtained in both interrogations.
	\end{proof}

	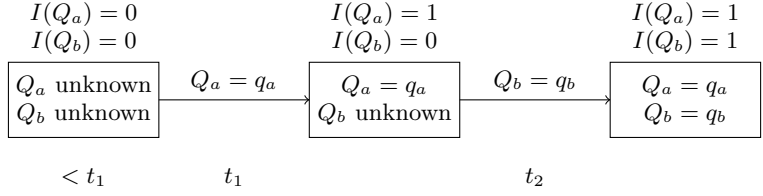
\begin{figure}[!h]
		\centering
		\begin{tikzpicture}
			
			\draw (1,2) rectangle (3,3) node[pos=.5, align=center] {$Q_a$ unknown\\$Q_b$ unknown};
			\node[above, align=center] at (2,3) {$I(Q_a)=0$\\$I(Q_b)=0$};
			
			\draw (5,2) rectangle (7,3) node[pos=.5,align=center] {$Q_a = q_a$\\$Q_b$ unknown};
			\node[above, align=center] at (6,3) {$I(Q_a)=1$\\$I(Q_b)=0$};
			
			\draw (9,2) rectangle (11,3) node[pos=.5,align=center] {$Q_a=q_a$\\$Q_b=q_b$};
			\node[above, align=center] at (10,3) {$I(Q_a)=1$\\$I(Q_b)=1$};
			
			\draw[->] (3,2.5) -- node[above] {$Q_a=q_a$} (5,2.5);
			\draw[->] (7,2.5) -- node[above] {$Q_b=q_b$} (9,2.5);
			
			\node[below, align=center] at (2,1.7) {$<t_1$};
			\node[below, align=center] at (4,1.7) {$t_1$};
			\node[below, align=center] at (8,1.7) {$t_2$};
			
		\end{tikzpicture}
		\caption{\textit{Mutually Compatible Questions in Interrogation} $Q_a$ and $Q_b$ are mutually compatible. Initially, we have no knowledge about the state and two questions. At time $t_1$, we conduct an interrogation of $Q_a$ with an outcome $q_a$. After this interrogation, we gain 1 unit information of $Q_a$ and cannot obtain any information about $Q_b$. At time $t_2$, we conduct another interrogation of $Q_b$ with outcome $q_b$, and information of $Q_a$ is reserved.}
	\end{figure}

	Based on the information of questions, we can then try to define the information of the system. Since we assume that the outcome probabilities of questions in $\mathcal{Q}_M$ determine all other outcome probabilities of questions in the set $\mathcal{Q}$, it is intuitive to define the information of the system as a sum of the information of questions in $\mathcal{Q}_M$, say~$\sum_{Q \in \mathcal{Q}_M} I(Q|h_{<t})$.
	
	However, simply taking the summation over $\mathcal{Q}_M$ may not be very useful, especially when dealing with a composite system, and $\mathcal{Q}_M$ contains compatible questions. Interrogations on compatible questions will retain each other's information, and they may also 'derive' information about non-interrogated questions.
	
	Consider a two-body qubit system where~$\{\hat{\sigma}_x \otimes \hat{\sigma}_x, \hat{\sigma}_y \otimes \hat{\sigma}_y, \hat{\sigma}_z \otimes \hat{\sigma}_z\}$ is a set of pairwise commuting operators. Under certain situations, say~$h_{\leq t_2}=\{``\hat{\sigma}_x \otimes \hat{\sigma}_x,+1,t_1",``\hat{\sigma}_z \otimes \hat{\sigma}_z,+1,t_2"\}$, from these two interrogations we may have 1 unit information for both~$\hat{\sigma}_x \otimes \hat{\sigma}_x$ and~$\hat{\sigma}_z \otimes \hat{\sigma}_z$. Moreover, now the system is in the common eigenstate of~$\hat{\sigma}_x \otimes \hat{\sigma}_x$ and~$\hat{\sigma}_z \otimes \hat{\sigma}_z$, which is just the Bell state~$\ket{\Psi^+}$, and the outcome probabilities of~$\hat{\sigma}_y \otimes \hat{\sigma}_y$ will be ensured, even if we haven't conducted an interrogation on it,
	\begin{equation}
		P(``\hat{\sigma}_y \otimes \hat{\sigma}_y,1,t_3"|h{\leq t_2},I) = 1, \quad P(``\hat{\sigma}_y \otimes \hat{\sigma}_y,0,t_3"|h{\leq t_2},I) = 0.
	\end{equation}
	
	This suggests that the information of~$\hat{\sigma}_y \otimes \hat{\sigma}_y$ is not independent, but can be derived from the results of the other two interrogations.
	
	\begin{figure}[!h]
		\centering
		\begin{tikzpicture}
			
			\draw (0.5,2) rectangle (3,3) node[pos=.5, align=center] {$\hat{\rho} = \frac{1}{4}\hat{I}$};
			\node[above, align=center] at (1.75,3) {$I(\s{x}\otimes\s{x})=0$\\$I(\s{y}\otimes\s{y})=0$\\$I(\s{z}\otimes\s{z})=0$};
			
			\draw (5.5,2) rectangle (8,3) node[pos=.5,align=center] {\scriptsize$\ket{\psi}\in E(\s{x}\otimes\s{x},1)$};
			\node[above, align=center] at (6.75,3) {$I(\s{x}\otimes\s{x})=1$\\$I(\s{y}\otimes\s{y})=0$\\$I(\s{z}\otimes\s{z})=0$};
			
			\draw (10.5,2) rectangle (13,3) node[pos=.5,align=center] {$\ket{\psi}=\ket{\Psi^+}$};
			\node[above, align=center] at (11.75,3) {$I(\s{x}\otimes\s{x})=1$\\$I(\s{y}\otimes\s{y})=1$\\$I(\s{z}\otimes\s{z})=1$};
			
			\draw[->] (3,2.5) -- node[above] {$\s{x}\otimes\s{x}=1$} (5.5,2.5);
			\draw[->] (8,2.5) -- node[above] {$\s{z}\otimes\s{z}=1$} (10.5,2.5);
			
			\node[below, align=center] at (1.75,1.7) {$<t_1$};
			\node[below, align=center] at (4.25,1.7) {$t_1$};
			\node[below, align=center] at (9.25,1.7) {$t_2$};
			
		\end{tikzpicture}
		\caption{\textit{Information Acquisition in Joint Measurements}
		Initially, we have no information about these joint measurements, and the initial state is taken as the maximal mixed state with no preference for any projection. At time $t_1$, a measurement of $\s{x}\otimes\s{x}$ is taken, projecting the system into the eigensubspace of $\s{x}\otimes\s{x}$. At time $t_2$, another measurement of $\s{z}\otimes\s{z}$ is conducted, ensuring that the system is in one of the four Bell states. All the information about the three joint measurements is obtained.
		}
	\end{figure}
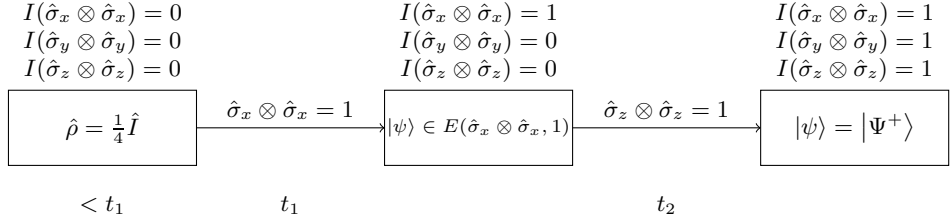
	
	
	
	We want to focus on those non-derived information only, the information of the system will be sum of these independent information in the subset~$\mathcal{Q}_M$: 
	\begin{equation}
		I_{system} (h_{<t}) = \sum_{\substack{Q_i \in \mathcal{Q}_M\\ Q_i ~rel.~ind.}} I(Q_i|h_{<t}).
	\end{equation}
	Here~$Q_i ~rel.~ind.$ denotes to all the~ questions such that the outcome probabilities cannot be derived from other questions in~$\mathcal{Q}_M$. In the above example when~$h_{<=t_2} = \{``\hat{\sigma}_x \otimes \hat{\sigma}_x,+1,t_1",``\hat{\sigma}_z \otimes \hat{\sigma}_z,+1,t_2"\}$, then~$\hat{\sigma}_y \otimes \hat{\sigma}_y$ will be excluded in the sum and all other questions in~$\mathcal{Q}_M$ will be retained.

	\assumption{} For $n$-ary single system, the upper bound of information of system is 1 unit. For composite system composed of~$N$ subsystems, this upper bound is~$N$ units.
	
	This assumption together with logical gates directly leads to the following corollaries:
	
	\cor{} In single~$n$-ary system, all questions in~$\mathcal{Q}_M$ are mutually complementary.
	
	\begin{proof}
		Let~$Q_a, Q_b \in \mathcal{Q}_M$, and we take interrogation on~$Q_a$ at time~$t_1$ with outcome~$q_a$ and interrogation on~$Q_b$ at time~$t_2$ with outcome~$q_b$.
		
		After~$t_2$ we may obtain 1 unit information about question~$Q_b$:
		\begin{equation}
			P(``Q_b,q'_b,t_3"|``Q_b,q_b,t_2",``Q_a,q_a,t_1",I)=\delta_{q'_b,q_b},\quad I(Q_b|``Q_b,q_b,t_2",``Q_a,q_a,t_1",I) = 1.
		\end{equation}
		
		The information of the system is no more than 1 unit:
		\begin{equation}
			I_{system} (``Q_b,q_b,t_2",``Q_a,q_a,t_1") = I(Q_a|``Q_b,q_b,t_2",``Q_a,q_a,t_1") + I(Q_b|``Q_b,q_b,t_2",``Q_a,q_a,t_1") \le 1.
		\end{equation}
		
		This yields~$I(Q_a|``Q_b,q_b,t_2",``Q_a,q_a,t_1") = 0$ and~$P(``Q_a,q'_a,t_3"|``Q_b,q_b,t_2",``Q_a,q_a,t_1",I)=\frac{1}{n}, \forall q'_a$
		
		By applying the same argument, if we first take interrogation of~$Q_b$ and then take interrogation of~$Q_a$ we may have~$P(``Q_b,q'_b,t_3"|``Q_a,q_a,t_2",``Q_b,q_b,t_1",I)=\frac{1}{n}~~ \forall q'_b $.
		
		The above procedures show that~$Q_a$ and~$Q_b$ are pairwise complementary, and choice of~$Q_a$ and~$Q_b$ are arbitrary. Therefore all questions in~$\mathcal{Q}_M$ are mutually complementary.
	\end{proof}
	
	\cor{} Two composite questions~$Q_a *_i Q_b, Q_{a'} *_j Q_{b'}$ are not compatible if~$Q_a = Q_{a'}$ or~$Q_b = Q_{b'}$, $*_i,*_j$ are any two allowable logical gates.
	
	\begin{proof}
		By contradiction, assume if~$Q_a = Q_{a'}$ then~$Q_a *_i Q_b, Q_{a} *_j Q_{b'}$ are compatible.
		
		According to Assumption 4, the three questions~$Q_a, Q_a *_i Q_b, Q_{a} *_j Q_{b'}$ are mutually compatible. This means we can make three interrogations on each of them and information on each question won't be lost.
		
		The outcomes of~$Q_a, Q_a *_i Q_b$ will yield the outcome of~$Q_b$ and outcomes of~$Q_a, Q_a *_j Q_{b'}$ will yield outcome of~$Q_{b'}$. Yet if~$Q_b, Q_{b'} \in \mathcal{Q}_{M_B}$ then this violate Corollary 3 since we cannot know the outcomes of two complementary questions.
		
		The same argument can be applied on the case that~$Q_b = Q_{b'}$.
	\end{proof}
	
	\cor{} In two body~$p$-ary system, there are at most~$p+1$ mutually compatible composite questions, where every composite question is in the form~$Q_a *_i Q_b$,~$Q_a \in \mathcal{Q}_{M_A}, Q_b \in \mathcal{Q}_{M_B}, *_i$ is any allowable logical gate.
	
	\begin{proof}
		Assume there are~$k$ different mutually compatible composite questions, which are labeled as~$Q_1 *_{i_1} Q_{j_1}, Q_2 *_{i_2} Q_{j_2},\cdots,Q_k *_{i_k} Q_{j_k}$.
		
		All those questions are compatible means we may take~$k$ those different interrogations, say~$h_{<=t_m}=\{``Q_m *_{i_m} Q_{j_m}, q_m, t_m"\}^k_{m=1}$, and after that information of each of the~$k$ different questions is retained.
		
		
		Yet for two body system we can only obtain at most 2 units information of system. Of course it's possible to require~$k\le2$ but it could be very trivial. We want to attain the maximal value of~$k$. If~$k>2$, even if information of every question is retained,
		~the information of system is still 2 units. 
		
		Assume we first take~$2$ interrogations,~$\{``Q_2 *_{i_2} Q_{j_2},q_2,t_2",``Q_1 *_{i_1} Q_{j_1},q_1,t_1"\}$, we will have 1 unit information for each of the question. Those two questions are independent to each other, this suggests that we must have 2 units information of the system.
		Yet if we take another~$k-2$ consecutive interrogations,~$\{``Q_n *_{i_m} Q_{j_m},q_m,t_m"\}_{m=1}^{k}$, the information of the system is still 2 units, it won't violate the upper bound,
		\begin{equation}
			I_{system}(h_{<=t_m}) = I_{system}({``Q_2 *_{i_2} Q_{j_2},q_2,t_2",``Q_1 *_{i_1} Q_{j_1},q_1,t_1"}) = 2.
		\end{equation}
		All the~$k$ questions are mutually compatible, and we could ensure their outcomes simultaneously.
		This means the remained~$k-2$ interrogations are pre-determined, the outcomes of those~$k-2$ interrogations must be determined from the outcomes of first~$2$ interrogations:
		$$\{q_3,q_4,\cdots,q_k\}\text{~are determined by~}\{q_1,q_2\}.$$
		In other words,~$\forall m\in \{3,4,\cdots,k\}\: \exists f_m :\mathbb{F}_p \times \mathbb{F}_p\rightarrow \mathbb{F}_p $ s.t.~$q_m = f_m(q_1, q_2) $. And there are two restrictions of function~$f_m$:
		\begin{enumerate}
			\item $q_m$ is independent with~$q_1$ and~$q_m$ is independent with~~$q_2$;
			\item $q_m$ is independent with~$q_{m'}$ if~$m\ne m'$.
		\end{enumerate}
		Those two restrictions come from the fact that~$\{Q_m *_{i_m} Q_{j_m}\}_{m=1}^{k} \subset \mathcal{Q}_{M_{AB}}$ and they are mutually independent, so are their outcomes. The domain and image of~$f_m$ are both collection of discrete numbers and it is possible to write down a truth table of~$f_m$. According to the discussion of logical gates above, in such a~$p$-ary system, there are only at most~$p-1$ such functions, which means~$k-2 \le p-1$. Therefore the maximal possible number of~$k$ is~$p+1$.

	\end{proof}
	
	\assumption{} In a two-body composite system, given two composite questions in the form~$Q_i *_n Q_j$ and~$Q_k *_m Q_l$, if~$i \ne k$ and~$j \ne l$, then for every logical gate~$*_n$, there exists a unique logical gate~$*_m$ such that~$Q_i *_n Q_j$ and~$Q_k *_m Q_l$ are compatible.
	
	This assumption is not very intuitive. From the viewpoint of tomography, we may regard each composite question as a combination of two questions asked on individual systems, and different logical gates shouldn't be affected much. However, this assumption actually arises from a fact of correspondence in linear space. Later, when discussing the quantum mechanical correspondence of the quantum question structure, we will provide a detailed proof of that fact. With the help of this assumption and Corollary 6, we will derive an important result.
	
	\theorem{} For single~$p$-ary system, the size of~$\mathcal{Q}_M$ is no more than~$p+1$.
	\begin{proof}
		By contradiction, assume size of~$\mathcal{Q}_M$ is large than~$p+1$, say equal to~$p+2$.
		
		Consider a two body~$p$-ary system. According to Corollary 5, there are at most~$p+1$ mutually compatible composite questions, and we may choose a set of them labeled as~$Q_1 *_{i_1} Q_{j_1}, Q_2 *_{i_2} Q_{j_2},\cdots,Q_{p+1} *_{i_{p+1}} Q_{j_{p+1}}$. 
		
		Let~$Q_{j_{p+2}} \in \mathcal{Q}_M \setminus \{Q_{j_1},Q_{j_2},\cdots,Q_{j_{p+1}} \}$, then~$Q_{p+2} *' Q_{j_{p+1}}$ is not in the collection of those~$(p+1)$ mutually commuting operators for any logical gate~$*'$.
		
		From Assumption 6, for every~$k \in \{1,2,\cdots,p+1\}$, there exists a unique logical gate~$*_{m_k}$ such that~$Q_k *_{i_k} Q_{j_k}$ and~$Q_{p+2} *_{m_k} Q_{j_{p+2}}$ are compatible.
		
		Since there are at most~$p-1$ different logical gates, this means there must be repetition among the collection of logical gates~$\{*_{m_1},*_{m_2},\cdots,*_{m_{p+1}}\}$. Let the repetition logical gate be~$*_{m_\alpha}=*_{m_\beta}$. Therefore~$Q_\alpha *_{i_\alpha} Q_{j_\alpha}$ and~$Q_\beta *_{i_\beta} Q_{j_\beta}$ are both compatible to~$Q_{p+2} *_{m_\alpha} Q_{j_{p+2}}$. 
		
		Follow from the argument of Corollary 5, $Q_{p+2} *_{m_\alpha} Q_{j_{p+2}}$ must be a function of~$Q_\alpha *_{i_\alpha} Q_{j_\alpha}$ and~$Q_\beta *_{i_\beta} Q_{j_\beta}$. Yet in the collection~$\{Q_1 *_{i_1} Q_{j_1}, Q_2 *_{i_2} Q_{j_2},\cdots,Q_{p+1} *_{i_{p+1}} Q_{j_{p+1}}\}$ all operators other than~$Q_\alpha *_{i_\alpha} Q_{j_\alpha}$ and~$Q_\beta *_{i_\beta} Q_{j_\beta}$ are also different functions of~$Q_\alpha *_{i_\alpha} Q_{j_\alpha}$ and~$Q_\beta *_{i_\beta} Q_{j_\beta}$. This suggests that~$Q_{p+2} *_{m_\alpha} Q_{j_{p+2}}$ will also be compatible with the collection of composite questions. 
		
		However,~$Q_{p+2} \notin \{Q_1,Q_2,\cdots,Q_{p+1}\} $ and~$Q_{j_{p+2}} \notin \{Q_{j_1},Q_{j_2},\cdots,Q_{j_{p+1}} \} $, $Q_{p+2} *_{m_\alpha} Q_{j_{p+2}}$ is different with any member of the collection~$\{Q_{l} *_{i_l} Q_{j_l}\}^{p+1}_{l=1}$. Now there are~$(p+2)$ different compatible composite questions, which contradicts with Corollary 5.
		
	\end{proof}
	
	\section{Correspondences in quantum mechanics}
	\label{section_correspondence}

	In the above discussions, the key concepts are compatible and complementary  questions. The former corresponds to commuting operators while the latter are closely related to the concept of mutually unbiasedness~\cite{Klimov2009, durt2010} in quantum mechanics.

	\definition{} Two non-degenerate operators~$\hat{A}$ and~$\hat{B}$ with~$d$ distinct eigenvalues are said to be \textit{mutually unbiased} if there is a set of orthonormal eigenstates~$\{\ket{a_n}\}$ of~$\hat{A}$ and a set of orthonormal eigenstates~$\{\ket{b_n}\}$ of~$\hat{B}$ such that
	\begin{equation}
		\left|\bra{a_i}\ket{b_j}\right|^2=\frac{1}{d} \qquad \forall i,j \in \mathbb{F}_d.
	\end{equation}
	
	The mutually unbiasedness between two non-degenerate operators is in fact determined by their eigenstates.
	The concept of mutually unbiased operators can be extended to degenerate operators where the two set of eigensubspaces are mutually unbiased.
	\definition{} Two degenerate operators~$\hat{A}$ and~$\hat{B}$ with~$d$ distinct eigenvalues~$\{\lambda_{Ai}\}_{i=1}^d, \{\lambda_{Bi}\}_{i=1}^d$ are said to be~\textit{mutually unbiased} if
	\begin{equation}
		\left|\bra{a_i}\ket{b_j}\right|^2=\frac{1}{d} \qquad \forall i,j \in \mathbb{F}_d \quad \ket{a_i}\in E(\lambda_{Ai},\hat{A})\quad \ket{b_j}\in E(\lambda_{Bj},\hat{B}).
	\end{equation}
	
	
	\definition{} In~$\mathbb{C}^d$, two orthonormal bases~$\{\ket{a_n}\},\{\ket{b_n}\}(n\in \{0,1,2,\cdots,d-1\})$ are said to be \textit{mutually unbiased bases}  if
	\begin{equation}
		\left|\bra{a_i}\ket{b_j}\right|^2=\frac{1}{d} \quad \forall i,j \in \mathbb{F}_d.
	\end{equation}
	In~$\mathbb{C}^d$, we can always find a set of orthonormal basis~$\{\ket{i}\} (i\in \{0,1,2,\cdots,d-1\})$ as \textit{computational basis}. Another set of orthonormal basis~$\{\ket{\tilde{j}}\}(j\in \{0,1,2,\cdots,d-1\})$ can be defined by quantum discrete Fourier transformation:
	\begin{equation}
	\ket{\tilde{j}}=\frac{1}{\sqrt{d}}\sum_{k=0}^{d-1}\omega_d^{-kj}\ket{k} \qquad \omega_d=e^{2i\pi/d },
	\end{equation}
	where $\{\ket{i}\}$ and~$\{\ket{\tilde{j}}\}$ are unbiased since~$\bra{i}\ket{\tilde{j}}=\frac{1}{\sqrt{d}}\omega_d^{-ij}$.
	
	Based on those two set of mutually unbiased bases, we can introduce \textit{generalized Pauli matrix} $\hat{X}$ and~$\hat{Z}$:
	\begin{equation}
	\hat{X}\ket{\tilde{j}}=\omega_d^j\ket{\tilde{j}},\qquad \hat{Z}\ket{i}=\omega_d^i\ket{i}.
	\end{equation}
	
	From the definition, it follows that~$\hat{X},\hat{Z}$ have the following important properties:
	\begin{enumerate}
		\item $\hat{X}^d=\hat{Z}^d=\hat{I}$;
		\item $\hat{X}\ket{i}=\ket{i+1}$,~$\bra{\tilde{j}}\hat{Z}=\bra{\widetilde{j+1}}$;
		\item $\hat{Z}\hat{X}=\omega_d\hat{X}\hat{Z}$ (Weyl commutation relation).
	\end{enumerate}
	
	\begin{table}[!h]
		\centering	
		$\hat{X}= \begin{bmatrix}
		0 & 0 & 0 & 0 & 1 \\
		1 & 0 & 0 & 0 & 0 \\
		0 & 1 & 0 & 0 & 0 \\
		0 & 0 & 1 & 0 & 0 \\
		0 & 0 & 0 & 1 & 0 
		\end{bmatrix}  \quad$
		$\hat{Z}= \begin{bmatrix}
		1 & 0 & 0 & 0 & 0 \\
		0 & \omega_5 & 0 & 0 & 0 \\
		0 & 0 & \omega^2_5 & 0 & 0 \\
		0 & 0 & 0 & \omega^3_5 & 0 \\
		0 & 0 & 0 & 0 & \omega^4_5 
		\end{bmatrix}  $
		\caption{Matrix representation of~$\hat{X},\hat{Z}$ in the computational basis of dimension 5}
	\end{table}
	
	\fact{} In~$\mathbb{C}^d$, there are at least three mutually unbiased bases which are the eigenstates of~$\{\hat{X},\hat{Z},\hat{X}\hat{Z}\}$.
	If~$d$ is a prime number, $\mathbb{C}^d$ has the maximal number of of MUBs, which are the eigenstates of~$\{\hat{X},\hat{Z},\hat{X}\hat{Z},\hat{X}\hat{Z}^2,\cdots,\hat{X}\hat{Z}^{d-1}\}$\cite{durt2010}. The eigenstate of~$\hat{X}\hat{Z}^k$ is expressed as:
	\begin{equation}
	 	\ket{e_k^j} = \frac{1}{\sqrt{d}}\sum_{i=0}^{d-1}\omega^{-ij}\omega^{ki(i-1)/2}\ket{i},
	\end{equation}
	where~$\ket{i}$ is the eigenstate of~$\hat{Z}$ and~$\hat{X}\hat{Z}^k\ket{e_k^j} = \omega^j\ket{e_k^j}$. 
	\section{Connections between question set structure and quantum mechanics}
	\label{section_connection}
	\subsection{Single system}
	
	\subsubsection{Relations between~$p$-ary question set structure and quantum mechanics}
	In the following, we list some connections that we have established between quantum mechanics in~$\mathbb{C}^p$ space and a system represented as set of~$p$-ary questions.

	\begin{table*}[!h]
		\centering
		\begin{ruledtabular}
			\begin{tabular}{p{7.75cm}p{7.75cm}} 
				{$p$-ary question set structure } 			 & {Quantum mechanics in~$\mathbb{C}^p$}   \\ \hline
				
				\smallskip Question $Q_a$ with~$p$ different outcomes
				& \smallskip{Unitary operator~$\hat{U}_a$ with~$p$ different eigenvalues}\\   
				
				\smallskip Questions in~$\mathcal{Q}_M$ are mutually complementary & \smallskip Bases of corresponding operators in~$\mathcal{Q}_M$ are MUBs \\
				
				\smallskip The probabilities of set~$\mathcal{Q}_M$ determines state of system
				& \smallskip Probabilities of projections of MUBs determines density matrix\\			
				
			\end{tabular}
		\end{ruledtabular}
		\caption{\textit{Comparison between Question Set Structure and Quantum Mechanics on Single System}}
	\end{table*}

	\begin{enumerate}
		\item Each question $A$ in the question set $\mathcal{Q}$ has $p$ different outcomes, which are ${0, 1, 2, \ldots, p-1}$. Question $A$ corresponds to a unitary operator $\hat{A}$ in $\mathbb{C}^p$ with $p$ distinct eigenvalues ${\omega^0_p, \omega^1_p, \omega^2_p, \ldots, \omega^{p-1}_p}$. While $\hat{A}$ is not Hermitian, we can always decompose it in terms of its eigenstate projectors, such as $\hat{A} = \sum_i \omega^{i-1}\ket{e_i}\bra{e_i}$. Every interrogation of question $Q_a$ corresponds to a collection of $p$ projections, with each projection onto an eigenstate of $\hat{A}$.	
		Of course, we can find a set of $p$ real values ${\lambda_0, \lambda_1, \ldots, \lambda_{p-1}}$ to create a Hermitian operator, denoted as $\hat{A}_{\text{Her.}} = \sum_i \lambda_i\ket{e_i}\bra{e_i}$, where $\lambda_i \in \mathbb{R}$. This makes it more natural to connect the interrogation of $Q_a$ to the measurement of $\hat{A}_{\text{Her.}}$. However, such a $\hat{A}_{\text{Her.}}$ is not convenient for the following calculations, especially for a composite system. Therefore, we will continue to use the interrogation-projection connection.
		\item The set $\mathcal{Q}_M$ corresponds to a maximal set of Mutually Unbiased Bases (MUBs) of $\mathbb{C}^p$. For each set of bases ${e_n}$ in a maximal set of MUBs, we can define a unitary or Hermitian operator based on ${e_n}$. In fact, the generalized Pauli matrices introduced above are used to label different bases in a set of MUBs.
	\end{enumerate}
	
	\subsubsection{Example of information changed on different interrogations}
	In a single~$p$-ary system, there are~$p+1$ questions in~$\mathcal{Q}_M$, say~$\mathcal{Q}_M=\{Q_1,Q_2,\cdots,Q_{p+1}\}$. Their corresponding operators are just the generalized Pauli matrices in~$\mathbb{C}^p$, $\{\hat{X},\hat{Z},\hat{X}\hat{Z},\hat{X}\hat{Z}^2,\cdots,\hat{X}\hat{Z}^{p-1}\}$. We will show how the information of single question
	
	\emph{Quantum Question Scenario}
	
	Assume we are facing an unknown system, all the knowledge we have is that this is a~$p$-ary outcome system. In this situation we may initialize the system being the state such that
	all the outcome probabilities of those questions are the same,
	\begin{equation}
		P(``Q_i,q_i,t_1"|h_{<t_1}=\emptyset,I)=\frac{1}{p} \quad \forall Q_i \in \mathcal{Q}_M\: \forall q_i \in \mathbb{F}_p,
	\end{equation}
	where~$h_{<t_1}=\emptyset$ denotes we know nothing about the system before time~$t_1$.

	In other words, the information of any question is zero unit, the information of system is also zero,
	\begin{equation}
		I(Q_i|h_{<t_1}) = 0~~ \forall Q_i \in \mathcal{Q}_M, \qquad I_{system}(h_{<t_1}) = 0.
	\end{equation}

	At time~$t_1$ we conduct an interrogation of~$Q_1$ and obtain an outcome~$m$. After this interrogation we obtain~$1$ unit information about~$Q_1$:
	\begin{equation}
		P(``Q_1,m',t_2"|``Q_1,m,t_1",h_{<t_1},I) = \delta_{m,m'} \quad I(Q_1|``Q_1,m,t_1",h_{<t_1}) = 1.
	\end{equation}
	All questions in~$\mathcal{Q}_M$ are mutually complementary, hence the outcome probabilities of all other questions are remained as uniform distributions,
	\begin{equation}
		\begin{aligned}
			P(``Q_i,q_i,t_2"|``Q_1,m,t_1",I) = \frac{1}{p}\quad &\forall Q_i \in \mathcal{Q}_M \: Q_i \ne Q_1 \: \forall q_i \in \mathbb{F}_p, \\
			I(Q_i|``Q_1,m,t_1") = 0\quad &\forall Q_i \in \mathcal{Q}_M \: Q_i \ne Q_1.
		\end{aligned}		
	\end{equation}

	The information of all other questions are just zero, when calculating the information of the system, we needn't to exclude any questions since there is only one non-zero term. The information of system is just one unit:
	\begin{equation}
		I_{system}(``Q_1,m,t_1",h_{<t_1}) = \sum_{Q_i \in \mathcal{Q}_M} I(Q_i|``Q_1,m,t_1",h_{<t_1}) = 1.
	\end{equation}
	
	At time~$t_2$ we take another interrogation of~$Q_2$ with outcome~$n$. After this interrogation we obtain 1 unit information about~$Q_2$:
	\begin{equation}
		P(``Q_2,n',t_3"|``Q_2,n,t_2",``Q_1,m,t_1",h_{<t_1},I) = \delta_{n,n'}, \quad I(Q_2|``Q_2,n,t_2",``Q_1,m,t_1",h_{<t_1},I) = 1.
	\end{equation}
	The information of~$Q_1$ is lost according to the second interrogation, and information of all other questions are still zero:
	\begin{equation}
		\begin{aligned}
			P(``Q_i,q_i,t_3"|``Q_2,n,t_2",``Q_1,m,t_1",h_{<t_1},I) = \frac{1}{p}\quad &\forall Q_i \in \mathcal{Q}_M~s.t.\: Q_i \ne Q_2 \: \forall q_i \in \mathbb{F}_p, \\
			I(Q_i|``Q_2,n,t_2",``Q_1,m,t_1",h_{<t_1}) = 0\quad &\forall Q_i \in \mathcal{Q}_M~s.t.\: Q_i \ne Q_2.
		\end{aligned}
	\end{equation}

	Similarly the information of the system is still~$1$ unit, it didn't exceed the upper bound,
	\begin{equation}
		I_{system}(``Q_2,n,t_2",``Q_1,m,t_1",h_{<t_1}) = \sum_{Q_i \in \mathcal{Q}_M} I(Q_i|``Q_2,n,t_2",``Q_1,m,t_1",h_{<t_1}) = 1
	\end{equation}
	
	\emph{Quantum Mechanics Scenario}
	
	At time~$t_1$, we take measurement~$\hat{X}$ on a single~$p$-dimensional system with an outcome~$\omega_p^m$. After this measurement, the state of the system is ensured, which is the eigenstate of~$\hat{X}$:
	\begin{equation}
		\ket{\psi}_{t>t_1} = \ket{\tilde{m}}.
	\end{equation}

	We gain~$1$ unit information about~$\hat{X}$ and zero information about all other measurements,
	\begin{equation}
		P(``\hat{X},\omega_p^{m'},t_2"|``\hat{X},\omega_p^{m},t_1",I) = \delta_{m,m'}, \quad I(\hat{X}|``\hat{X},\omega_p^{m},t_1") = 1.
	\end{equation}
	
	The information of the system is~$1$ unit, achieving the upper bound. As expected, when information of the system achieving the upper bound we could ensure the state.
	
	At time~$t_2$, we take another measurement~$\hat{Z}$ with an outcome~$\omega_p^n$. After the second measurement, the state of the system is now changed to the eigenstate of~$\hat{Z}$:
	\begin{equation}
		\ket{\psi}_{t>t_2} = \ket{{n}}.
	\end{equation}

	We now gain~$1$ unit information about~$\hat{Z}$ and the information of~$\hat{X}$ is lost. Information of all other measurements in~$\mathcal{Q}_M$ remain the same. The information of the system is still~$1$ unit.

	\subsubsection{Interpretation of size of~$\mathcal{Q}_M$}
	The density matrix of a quantum system, which lies in $\mathbb{C}^p$, has a degree of freedom of $p^2-1$ due to hermiticity and normalization. In the question set structure, every question in $\mathcal{Q}_M$ can be represented as a $p$-tuple probability distribution, resulting in every question having $p-1$ degrees of freedom. As deduced above, for a single $p$-ary question set system, there are a total of $p+1$ mutually independent questions, and the total degree of freedom of $\mathcal{Q}_M$ will be exactly $(p-1)(p+1) = p^2-1$, which is equal to the degree of freedom of the density matrix of a $p$-dimensional system. This shouldn't be surprising, as each value in a $p$-tuple probability distribution corresponds to the probability of a projection.
	
	This is consistent with our assumption that $\mathcal{Q}_M$ contains the smallest number of $p$-outcome questions that determine the state of the system, as well as the probability distributions of questions in the set $\mathcal{Q}\setminus\mathcal{Q}_M$.
	
	\subsection{Two-body composite system}
	\subsubsection{Additional relations for composite systems}
	Based on the connections above and the discussion of logical gates for a $p$-ary question set structure, we propose a set of connections between a two-body $p$-ary system under the question set structure and quantum mechanics in $\mathbb{C}^p \otimes \mathbb{C}^p$.

	\begin{table*}[!h]
		\centering
		\begin{ruledtabular}
			\begin{tabular}{p{7.75cm}p{7.75cm}} 
				{$p$-ary two-body system in quantum question structure} 			 & Quantum mechanics in~$\mathbb{C}^p\otimes \mathbb{C}^p$   \\ \hline
				
				\smallskip Question $Q_a$ with~$p$ different outcomes
				& \smallskip Unitary operator~$\hat{U}_a$ with~$p$ different eigenvalues\\   
				
				\smallskip The probabilities of set~$\mathcal{Q}_M$ determines state of system & \smallskip Probabilities of projections of MUBs determines density matrix \\
				
				\smallskip Composite question is in the form of~$Q_a *_i Q_b$
				& \smallskip Composite operator in the form of~$\hat{U}_a\otimes\hat{U}_b^i$\\			
				
			\end{tabular}
		\end{ruledtabular}
		\caption{\textit{Comparison between Question Set Structure and Quantum Mechanics on Two-body System}}
	\end{table*}
	
	\begin{enumerate}
		\item For a two-body composite system, composite question~$Q_a*_i Q_b$ is related to the composite operator~$\hat{U}_a\otimes\hat{U}_b^i$ in~$\mathcal{L}(\mathbb{C}^p\otimes\mathbb{C}^p)$. $\hat{U}_a\otimes\hat{U}_b^i$ has~$p$ distinct eigenvalues~$\{\omega_p^0,\omega_p^1,\cdots,\omega_p^{p-1}\}$, each eigenvalue has degeneracy~$p$.
		\item $\mathcal{Q}_M$ for composite system contains both complementary and compatible questions. If two questions are compatible then their corresponding unitary operators commute.
	\end{enumerate}
	
	\subsubsection{Example of information changed when interrogating compatible questions}
	Here is an example of interrogations on composite compatible questions on 5-dimension, consider a family of commuting operators, $\{\hat{X}\otimes\hat{X}, \hat{Z}\otimes\hat{Z}^4, \hat{X}\hat{Z}\otimes\hat{X}\hat{Z}^4, \hat{X}\hat{Z}^2\otimes\hat{X}\hat{Z}^3, \hat{X}\hat{Z}^3\otimes\hat{X}\hat{Z}^2, \hat{X}\hat{Z}^4\otimes\hat{X}\hat{Z}\}$, and label them in form of composite questions, $\{Q_1 *_1 Q_1, Q_2 *_4 Q_2, Q_3 *_1 Q_6, Q_4 *_1 Q_5, Q_5 *_1 Q_4, Q_6 *_1 Q_3 \}$.
	
	\emph{Quantum Question Scenario}
	
	At time~$t_1$ if we take an interrogation of question~$Q_1 *_1 Q_1$ and obtain an outcome~$m$, we could immediately write down the following two outcome distributions:
	\begin{equation}
		\begin{aligned}
			&P(``Q_1 *_1 Q_1,m',t_2"|``Q_1 *_1 Q_1,m,t_1",I) = \delta_{m,m'}, \\
			&P(``Q,q,t_2"|``Q_1 *_1 Q_1,m,t_1",I) = \frac{1}{5}~~~ \forall Q \ne Q_1 *_1 Q_1 ~~ \forall q \in \mathbb{F}_5.
		\end{aligned}		
	\end{equation}
	This suggests that we gain 1 unit information about~$Q_1 *_1 Q_1$,
	\begin{equation}
		I(Q_1 *_1 Q_1|``Q_1 *_1 Q_1,m,t_1") = 1,
	\end{equation}
	but 0 unit information of all other questions,
	\begin{equation}
		I(Q|``Q_1 *_1 Q_1,m,t_1") = 0~~~ \forall Q \ne Q_1 *_1 Q_1.
	\end{equation}
	The information of the system is 1 unit:
	\begin{equation}
		\begin{aligned}
			I_{system}(``Q_1 *_1 Q_1,m,t_1") &= \sum_{Q \in \mathcal{Q}_M} I(Q|``Q_1 *_1 Q_1,m,t_1") \\
			&= I(Q_1 *_1 Q_1|``Q_1 *_1 Q_1,m,t_1") + \sum_{\substack{Q \in \mathcal{Q}_M\\
			Q \ne Q_1 *_1 Q_1
			}} I(Q|``Q_1 *_1 Q_1,m,t_1") \\
			&=		1 + 0 + 0 + \cdots = 1.
		\end{aligned}		
	\end{equation}
	Here when calculating the information of the system, we could sum the information of all questions in~$\mathcal{Q}_M$ without excluding any questions. The only non-zero information we obtain is on question~$Q_1 *_1 Q_1$, and all other questions are independent with it, nothing else could be derived from the outcome of~$Q_1 *_1 Q_1$.
	
	We then take an interrogation of question~$Q_2 *_4 Q_2$ at time~$t_2$ and obtain an outcome~$n$,
	\begin{equation}
		P(``Q_2 *_4 Q_2,n',t_3"|``Q_2 *_4 Q_2,n,t_2",``Q_1 *_1 Q_1,m,t_1",I) = \delta_{n,n'}.
	\end{equation}
	After the second interrogation, we got 1 unit information about~$Q_2 *_4 Q_2$:
	\begin{equation}
		I(Q_2 *_4 Q_2|``Q_2 *_4 Q_2,n,t_2",``Q_1 *_1 Q_1,m,t_1") = 1.
	\end{equation}
	The information of~$Q_1 *_1 Q_1$ is not~``lost", since~$Q_1 *_1 Q_1$ is compatible with~$Q_2 *_4 Q_2$,
	\begin{equation}
		\begin{aligned}
			P(``Q_1 *_1 Q_1,m',t_3"|``Q_2 *_4 Q_2,n,t_2",``Q_1 *_1 Q_1,m,t_1",I) = \delta_{m,m'}, \\
			I(Q_1 *_1 Q_1|``Q_2 *_4 Q_2,n,t_2",``Q_1 *_1 Q_1,m,t_1") = 1.
		\end{aligned}
	\end{equation}

	The information of~$Q_1 *_1 Q_1$ and~$Q_2 *_4 Q_2$ are independent to each other, the outcomes of both questions are not related. Therefore the information of the system must contain at least these two questions:
	\begin{equation}
		\begin{aligned}
			I_{system}(``Q_2 *_4 Q_2,n,t_2",``Q_1 *_1 Q_1,m,t_1") \ge I(Q_1 *_1 Q_1|``Q_2 *_4 Q_2,n,t_2",``Q_1 *_1 Q_1,m,t_1") + \\
			I(Q_2 *_4 Q_2|``Q_2 *_4 Q_2,n,t_2",``Q_1 *_1 Q_1,m,t_1") =2.
		\end{aligned}
	\end{equation}
	
	The information of the system hits the upper bound, 2 units. Even if we can still gain information about other four compatible composite questions,~$\{Q_3 *_1 Q_6, Q_4 *_1 Q_5, Q_5 *_1 Q_4, Q_6 *_1 Q_3 \}$, they are not independent and can be derived from the information of~$Q_1 *_1 Q_1$ and~$Q_2 *_4 Q_2$. In fact from Corollary 5, the outcomes of the other four compatible composite questions must be a function of outcomes of~$Q_1 *_1 Q_1$ and~$Q_2 *_4 Q_2$:
	\begin{equation}
		\begin{aligned}
			Q_3 *_1 Q_6 =& (Q_1 *_1 Q_1) *_1 (Q_2 *_4 Q_2) = m *_1 n = m+n, \\
			Q_4 *_1 Q_5 =& (Q_1 *_1 Q_1) *_2 (Q_2 *_4 Q_2) = m *_2 n = m+ 2n,\\
			Q_5 *_1 Q_4 =& (Q_1 *_1 Q_1) *_3 (Q_2 *_4 Q_2) = m *_3 n = m+3n,\\
			Q_6 *_1 Q_3 =& (Q_1 *_1 Q_1) *_4 (Q_2 *_4 Q_2) = m*_4 n = m+4n.
		\end{aligned}
	\end{equation}
	
	\emph{Quantum Mechanics Mcenario}
	
	At time~$t_1$, we take measurement of~$\hat{X}\otimes\hat{X}$ and obtain an outcome~$\omega_5^m$. After this measurement, we cannot write down the state of system, since~$\hat{X}\otimes\hat{X}$ is degenerate and all we can know is that the state of system lies in the eigensubspace of~$\hat{X}\otimes\hat{X}$. The outcome probability of all other measurements remains unknown,
	\begin{equation}
		\ket{\psi}_{t>t_1} \in E(\omega_5^m, \hat{X}\otimes\hat{X}).
	\end{equation}

	We gain 1 unit information about~$\hat{X}\otimes\hat{X}$, but 0 unit information of all other measurements. The information of the system is 1 unit.
	
	We then take measurement~$\hat{Z}\otimes\hat{Z}^4$ with outcome~$\omega_5^n$ at time~$t_2$.  Based on these two measurements, we can then write down the state of the system, which is a maximally entangled state:
	\begin{equation}
		\ket{\psi}_{t>t_2} = \frac{1}{\sqrt{5}}\left(\ket{0}\ket{\frac{m}{4}}+\omega_5^{4n}\ket{1}\ket{\frac{m-1}{4}}+ \omega_5^{3n}\ket{2}\ket{\frac{m-2}{4}} + \omega_5^{2n}\ket{3}\ket{\frac{m-3}{4}} + \omega_5^n \ket{4}\ket{\frac{m-4}{4}} \right).
	\end{equation}

	After the second interrogation, we got 1 unit information about~$\hat{Z}\otimes\hat{Z}^4$. The information of~$\hat{X}\otimes\hat{X}$ is retained, since~$\hat{X}\otimes\hat{X}$ and~$\hat{Z}\otimes\hat{Z}^4$ commute. The outcome probability of the other four commuting composite operators,~$\{\hat{X}\hat{Z}\otimes\hat{X}\hat{Z}^4, \hat{X}\hat{Z}^2\otimes\hat{X}\hat{Z}^3, \hat{X}\hat{Z}^3\otimes\hat{X}\hat{Z}^2, \hat{X}\hat{Z}^4\otimes\hat{X}\hat{Z} \}$, are ensured:
	\begin{equation}
		\begin{aligned}
			\hat{X}\hat{Z}\otimes\hat{X}\hat{Z}^4 \ket{\psi}_{t>t_2} =& \omega_5^{m+n} \ket{\psi}, \\
			\hat{X}\hat{Z}^2\otimes\hat{X}\hat{Z}^3 \ket{\psi}_{t>t_2} =& \omega_5^{m+2n} \ket{\psi}, \\
			\hat{X}\hat{Z}^3\otimes\hat{X}\hat{Z}^2 \ket{\psi}_{t>t_2} =& \omega_5^{m+3n} \ket{\psi}, \\
			\hat{X}\hat{Z}^4\otimes\hat{X}\hat{Z} \ket{\psi}_{t>t_2} =& \omega_5^{m+4n} \ket{\psi} .
		\end{aligned}
	\end{equation}
	
	This means the information of those four operators are derived from the information of~$\hat{X}\otimes\hat{X}$ and~$\hat{Z}\otimes\hat{Z}^4$. When calculating the information of the system, we need to exclude those four operators. Moreover, based on this state, the outcome probability of all other non-commuting composite operators will be just the uniform distribution. The information of the system will be 2 units, as expected, since the state of the system is already ensured.

	\subsubsection{Size of~$\mathcal{Q}_M$ and degree of freedom of density matrix}
	For a single system, we find that the degree of freedom of $\mathcal{Q}_M$ is equal to the degree of freedom of the density matrix. In fact, this relation also holds for a two-body composite system. Since all composite questions also have $p$ outcomes, every question in $\mathcal{Q}_M$ can be represented as a $p$-tuple probability distribution with $p-1$ degrees of freedom.
	
	As for the size of~$\mathcal{Q}_M$, consider a composite system~$\mathcal{Q}_{AB}$ contains two individual system~$\mathbb{Q}_A,\mathbb{Q}_B$, the number of elements in~$\mathcal{Q}_{M_{AB
	}}$ can be derived from its structure. Recall that from assumption 2,~$\mathcal{Q}_{M_{AB}}=\mathcal{Q}_{M_{A}}\cup\mathcal{Q}_{M_{B}}\cup\tilde{\mathcal{Q}}_{M_{AB}}$ and~$\tilde{\mathcal{Q}}_{M_{AB}}=\{Q_a*_iQ_b|Q_a \in \mathcal{Q}_{M_{A}}, Q_b \in \mathcal{Q}_{M_{B}},i\in \mathbb{F}^*_p\}$, therefore we could calculate the cardinality of $\mathcal{Q}_{M_{AB}}$:
	\begin{equation}
	\begin{aligned}
	\left|\mathcal{Q}_{M_{AB}}\right|&=\left|\mathcal{Q}_{M_{A}}\right|+\left|\mathcal{Q}_{M_{B}}\right|+\left|\tilde{\mathcal{Q}}_{M_{AB}}\right| \\
	&=(p+1) + (p+1) + (p+1)(p+1)(p-1) \\
	&=(p+1)(p^2+1).
	\end{aligned}
	\end{equation}
	
	The total degree of freedom of $\mathcal{Q}_{M_{AB}}$ is now equal to the product of $\left|\mathcal{Q}_{M_{AB}}\right|$ and $p-1$, which is equal to $p^4-1$, the same number of degrees of freedom as the density matrix of a two-body $p$-dimensional system in quantum mechanics.
	
	The agreement of degrees of freedom can be generalized to an $N$-body $p$-ary/$p$-dimensional system. In this situation, $\mathcal{Q}_M$ will contain composite questions from single systems, two-body systems, and so on, up to $N$-body systems. The total number of questions in $\mathcal{Q}_M$ is given by:
	\begin{equation}
		\begin{aligned}
			\binom{N}{1}(p+1) + \binom{N}{2}(p+1)^2(p-1) + \binom{N}{3}(p+1)^3(p-1)^2 +\cdots +\binom{N}{N}(p+1)^N(p-1)^{N-1}
			= \: \frac{p^{2N} - 1}{p - 1}.
		\end{aligned}
	\end{equation}
	The total degree of freedom of $\mathcal{Q}_M$ is $p^{2N} - 1$, which is the same as the degree of freedom of the density matrix for an $N$-body $p$-dimensional quantum system.
	
	\section{Conclusion}
	
	\textbf{Summary}
	In this paper, we present a new way to describe finite-dimensional quantum systems without relying on the language of linear spaces.
	
	We begin with an ideal finite-dimensional quantum system, where every measurement yields the same and a finite number of outcomes. We can abstract each measurement as posing a question to the system. We introduce a new set of assumptions, motivated by quantum tomography and information theory, to deduce the relationships between these questions. Among these assumptions, we use classical logical gates to represent joint measurements in composite quantum systems. By rewriting the logical gate operations in terms of truth tables, we find that all feasible logical gates can be connected through a specific orthogonal array, which exhibits special properties in prime-number-dimensional cases.
	
	The information of a single question is used to quantify our knowledge of a particular question based on a given background. It is defined as the generic entropy of the probability distribution of all possible outcomes for that question. The information of the system is used to quantify our knowledge of the entire system based on a specific background. It is defined as the sum of the information from selected questions, where these selected questions are assumed to characterize the state of the whole system. The constraints on logical gates and the assumptions regarding the information of the system lead us to derive detailed relations among these selected questions. These selected questions are mutually complementary, akin to the complementarity between Pauli matrices. The number of selected questions is also determined, which is $p+1$ for a $p$-ary system. Furthermore, the degree of freedom of the system is found to be the same as the degree of freedom of the density matrix of the corresponding quantum system.
	
	We have also established a connection between this new structure and the conventional quantum system. This bridge is built on the concept of mutually unbiased bases, which share similarities in construction and properties with orthogonal arrays. Each concept we introduce in this new framework has a direct correspondence with concepts in the conventional quantum system. Furthermore, through derivations within this new structure, we have uncovered new insights into conventional quantum systems with the assistance of this connection.
	
	\textbf{Comparison to other people's work}
	As mentioned in the introduction, our work is built upon a lineage of research that originated with Rovelli and was continued by Brukner, Zeillinger, and H\"{o}hn. All of these works aim to employ information to describe quantum theory. There are two significant differences between our work and the work of others.
	The first difference is our emphasis on arbitrary prime-number-dimensional systems, rather than exclusively binary systems. The binary case, as we have demonstrated in the section on the representation of logical gates, is somewhat fortuitous.
	The second difference is our attempt to alleviate confusion regarding the definition of information, encompassing both the information of a question and the information of a system. Ideally, we seek to represent the information of a system as a combination of information from specific questions. It is essential to determine the types of questions required to characterize the system.
	In table~\ref{comparison_information} we show the comparison between our work and three selected work.
		
	\begin{table}[!h]		
		\centering
		\begin{ruledtabular}
			\begin{tabular}{p{3.5cm}p{2.5cm}p{2.5cm}p{2.5cm}p{2.5cm}} 
				\smallskip& \smallskip Rovelli                       & \smallskip Zeillinger\&Brukner                                    & \smallskip H\"{o}hn                                     & \smallskip Our work                                   \\ \hline
				
				\smallskip Basic carrier of unit information   & \smallskip Binary outcome question & \smallskip Binary outcome question               & \smallskip Binary outcome question               & \smallskip \textbf{p-ary} outcome question                               \\ 
				\smallskip Finiteness of information of system & \smallskip Yes                     & \smallskip Yes                                   & \smallskip Yes                                   & \smallskip Yes                                                  \\ 
				\smallskip Information of single system        & \smallskip Sum of complete questions                       & \smallskip Sum of all complementary questions    & \smallskip Sum of independent bits & \smallskip Sum of all questions in~$\mathcal{Q}_M$                \\ 
				\smallskip Information of composite system     &    \smallskip                      & \smallskip Sum of selected correlation questions & \smallskip Sum of independent bits & \smallskip Sum of \textbf{rel. ind. questions} in~$\mathcal{Q}_M$ compare to context\\				
			\end{tabular}
		\end{ruledtabular}
		\caption{\label{comparison_information}Comparison between the four ideas that describing information of system. (1)``p-ary outcome question" denotes to an arbitrary prime number dimensional system. We extend the type of discussed system from binary case to higher dimensional case.
		(2)~We provide a criterion to choose the selected questions for characterizing composite system, that is, the mutually independent questions and the choice will be changed under different background. 
		}
	\end{table}

	We would like to emphasize that the most crucial aspect of incorporating information theory into quantum systems is to clarify the notion and definition of the concept of information, as well as the interpretation of probability distributions. Since most information measures are based on probabilities, we contend that the majority of the confusions surrounding information stem from an unclear understanding of probability distributions.
	
	\textbf{Discussion on unsolved problems}
	In this paper, we define the information of a single measurement, similar to Shannon entropy, although we do not provide a detailed expression. At present, we have not identified a compelling reason to choose a specific measure, and the extreme case of this measure suffices for our purposes. We leave this measure in its general form, which may be useful for deriving other ideas. We also point out the challenge of defining the information of a system, for which we do not present a complete solution but suggest a way to improve it.
	
	It is worth noting that both logical gates and MUBs have special properties in prime-number dimensions. This is why we focus exclusively on prime-number dimensional systems and why we were motivated to establish a connection between the question formalism and conventional quantum mechanics. Such a coincidence may have a profound mathematical explanation. Unfortunately, dealing with non-prime number dimensional cases is considerably more challenging. The construction of orthogonal arrays and MUBs is mathematically difficult in arbitrary finite dimensions, and whether this difficulty has a physical significance remains to be observed.
	
	\bibliographystyle{unsrt}
	\bibliography{QuanQuesBib}

\begin{thebibliography}{10}

\bibitem{twoqubitsDM}
Jing-Ling Chen, Libin Fu, Abraham~A. Ungar, and Xian-Geng Zhao.
\newblock Degree of entanglement for two qubits.
\newblock {\em Phys. Rev. A}, 65:044303, Apr 2002.

\bibitem{Rovelli1996}
Carlo Rovelli.
\newblock Relational quantum mechanics.
\newblock {\em International Journal of Theoretical Physics}, 35:1637, 1996.

\bibitem{BruknerZeilinger2002}
Caslav Brukner and Anton Zeilinger.
\newblock Information and fundamental elements of the structure of quantum
  theory.
\newblock In {\em Time, Quantum, and Information}, 2002.

\bibitem{BruknerZeilinger2009}
Caslav Brukner and Anton Zeilinger.
\newblock Information invariance and quantum probabilities.
\newblock {\em Foundations of Physics}, 39:677, 2009.

\bibitem{Hohn2017}
Philipp~Andres H{\"{o}}hn.
\newblock Quantum theory from rules on information acquisition.
\newblock {\em Entropy}, 19(3), 2017.

\bibitem{Hohn2017toolbox}
Philipp~Andres H{\"{o}}hn.
\newblock Toolbox for reconstructing quantum theory from rules on information
  acquisition.
\newblock {\em {Quantum}}, 1:38, December 2017.

\bibitem{Rao2009}
C.~Radhakrishna Rao.
\newblock {O}rthogonal arrays.
\newblock {\em Scholarpedia}, 4(7):9076, 2009.
\newblock revision \#137077.

\bibitem{OA}
John~Stufken A.S.~Hedayat, N. J. A.~Sloane.
\newblock {\em Orthogonal Arrays: Theory and Applications}.
\newblock Springer series in statisties. Springer-Verlag New York, 1 edition,
  1999.

\bibitem{Klimov2009}
Andrei~B. Klimov, Denis Sych, Luis~L. S\'anchez-Soto, and Gerd Leuchs.
\newblock Mutually unbiased bases and generalized bell states.
\newblock {\em Phys. Rev. A}, 79:052101, May 2009.

\bibitem{durt2010}
T.~Durt, B.~Englert, I.~Bengtsson, and K.~Zyczkowski.
\newblock On mutually unbiased bases.
\newblock {\em International Journal of Quantum Information}, 08(04):535--640,
  2010.

\bibitem{quant-ph/0103162}
Somshubhro Bandyopadhyay, P.~Oscar Boykin, Vwani Roychowdhury, and Farrokh
  Vatan.
\newblock A new proof for the existence of mutually unbiased bases.
\newblock 2001.

\bibitem{inadequacyShannon}
\ifmmode \check{C}\else~\v{C}\fi{}aslav Brukner and Anton Zeilinger.
\newblock Conceptual inadequacy of the shannon information in quantum
  measurements.
\newblock {\em Phys. Rev. A}, 63:022113, Jan 2001.

\end{thebibliography}

	\newpage
	
	\appendix
	
	\newpage

	\section{Correspondence of Corollary 6 in quantum mechanics}
	\lem{} In~$\mathbb{C}^p\otimes\mathbb{C}^p$, $\forall$ composite operators in the form of~$\compop{A}{}{B}{}{k}$ where~$\hat{A}, \hat{B} \in \{\hat{X},\hat{Z},\hat{X}\hat{Z},\hat{X}\hat{Z}^2,\cdots,\hat{X}\hat{Z}^{p-1}\}$ and~$k \in \{1,2,\cdots,p-1\}$, there are at most~$(p+1)$ different mutually commuting composite operators.
	
	\begin{proof}

		By contradiction, assume there are at least~$(p+2)$ different mutually composite operators, say~$\compop{A}{1}{B}{1}{k_1}, \compop{A}{2}{B}{2}{k_2}, \cdots, \compop{A}{p+2}{B}{p+2}{k_{p+2}} $.
		
		Since the choice of~$\hat{B}_i$ is limited, there will be~$n$ and~$m$ such that~$1\le n < m \le p+1$ and~$\hat{B}_n = \hat{B}_m$. This leads to three situations:
		\begin{enumerate}
			\item If~$\hat{A}_n \ne \hat{A}_m$, then this leads to contradiction since~$\left[\compop{A}{n}{B}{n}{k_n}, \compop{A}{m}{B}{n}{k_m} \right] \ne 0 $.
			\begin{equation}
				\begin{aligned}
					\left[\compop{A}{n}{B}{n}{k_n}, \compop{A}{m}{B}{m}{k_m} \right] =& \hat{A}_n\hat{A}_m\otimes(\hat{B}_n)^{k_n}(\hat{B}_n)^{k_m} - \hat{A}_m\hat{A}_n\otimes(\hat{B}_n)^{k_m}(\hat{B}_n)^{k_n} \\
					=&\left(\hat{A}_n\hat{A}_m - \hat{A}_m\hat{A}_n\right)\otimes\hat{B}_n^{k_n+k_m}
				\end{aligned}
			\end{equation}
			Without loss of generality, assume~$\hat{A}_n = \hat{X}^{i_n}\hat{Z}^{j_n}, \hat{A}_m = \hat{X}^{i_m}\hat{Z}^{j_m}$, where~$i_n, i_m \in \{0,1\} ~~ j_n = \delta_{i_n, 0} + n\delta_{i_n,1} ~~ j_m = \delta_{i_m, 0} + m\delta_{i_m,1} ~~ n,m\in \mathbb{F}^*_p $.
			By using Weyl commutation relation, we may have the following result:
			\begin{equation}
				\begin{aligned}
					\hat{A}_n\hat{A}_m - \hat{A}_m\hat{A}_n =& \hat{X}^{i_n}\hat{Z}^{j_n}\hat{X}^{i_m}\hat{Z}^{j_m} - \hat{X}^{i_m}\hat{Z}^{j_m}\hat{X}^{i_n}\hat{Z}^{j_n} \\
					=&\hat{X}^{i_n}\hat{Z}^{j_n}\hat{X}^{i_m}\hat{Z}^{j_m} -\omega_p ^{ i_nj_m-i_mj_n} \hat{X}^{i_n}\hat{Z}^{j_n}\hat{X}^{i_m}\hat{Z}^{j_m}
				\end{aligned}
			\end{equation}
			However, $i_nj_m-i_mj_n \ne 0 ~~ (\text{mod } p)$ since~$\hat{A}_n \ne \hat{A}_m$. This means~$\omega_p ^{i_mj_n - i_nj_m} \ne 1 $ and~$\left[\compop{A}{n}{B}{n}{k_n}, \compop{A}{n}{B}{m}{k_m} \right] \ne 0$.
			\item If~$\hat{A}_n = \hat{A}_m$ and~$k_n \ne k_m$, then the common eigensubspace of these two operators is ensured. 
			Let~$\ket{a}$ be the eigenstate of~$\hat{A}_n$ and~$\hat{b}$ be the eigenstate of~$\hat{B}_n$ such that~$\hat{A}_n\ket{a} = \omega^{a}\ket{a}$ and~$\hat{B}_n\ket{b} = \omega^{b}\ket{b}$. The eigenspace of~$\compop{A}{n}{B}{n}{k_n}$ and~$\compop{A}{n}{B}{n}{k_m}$ can then be expressed in terms of~$\ket{a}\otimes\ket{b}$:
			\begin{equation}
				\begin{aligned}
					&E(\omega_p^n, \compop{A}{n}{B}{n}{k_n}) = Span(\ket{a_1}\otimes\ket{b_1}, \ket{a_2}\otimes\ket{b_2}, \cdots, \ket{a_p}\otimes\ket{b_p})\quad &&a_i + k_n b_i = n \\
					&E(\omega_p^m, \compop{A}{n}{B}{n}{k_m}) = Span(\ket{a_1}\otimes\ket{b_1}, \ket{a_2}\otimes\ket{b_2}, \cdots, \ket{a_p}\otimes\ket{b_p})\quad &&a_i + k_m b_i = m
				\end{aligned}
			\end{equation}
			The common eigenspace of~$\compop{A}{n}{B}{n}{k_n}$ and~$\compop{A}{n}{B}{n}{k_m}$ can then be determined.
			\begin{equation}
				E(\omega_p^n, \compop{A}{n}{B}{n}{k_n}) \cap E(\omega_p^m, \compop{A}{n}{B}{n}{k_m}) = span(\ket{a_{n,m}}\otimes\ket{b_{n,m}}) 
			\end{equation}
			where~$a_{n,m}+k_n b_{n,m} = n\: (\text{mod } p)$ and~$a_{n,m}+k_m b_{n,m} = m\: (\text{mod } p)$.
			
			$\forall \hat{A}_l , \hat{B}_l \in \{\hat{X},\hat{Z},\hat{X}\hat{Z},\hat{X}\hat{Z}^2,\cdots,\hat{X}\hat{Z}^{p-1}\}$, we have\cite{quant-ph/0103162}:
			\begin{equation}
				\begin{aligned}
					\hat{A}_l \ket{a_{n,m}} &= \ket{a_{n,m}\oplus a_l} \\
					\hat{B}_l \ket{b_{n,m}} &= \ket{b_{n,m}\oplus b_l}
				\end{aligned}
			\end{equation}
			where~$\oplus$ is the addition in~$\mathbb{F}_p$ and~$a_l, b_l \in \mathbb{F}_p$. $a_l = 0$ if and only if~$\hat{A}_l = \hat{A}_n $, $b_l = 0$ if and only if~$\hat{B}_l = \hat{B}_n $.
			\begin{equation}
				\compop{A}{l}{B}{l}{k_l} \ket{a_{n,m}}\otimes\ket{b_{n,m}} = \ket{a_{n,m}\oplus a_l}\otimes\ket{b_{n,m}\oplus k_l b_l}
			\end{equation}
			
			This suggests that~$\ket{a_{n,m}}\otimes\ket{b_{n,m}} $ cannot be the eigenstate of~$\compop{A}{l}{B}{l}{k_l}$ if~$\hat{A}_n \ne \hat{A}_l $ or~$\hat{B}_n \ne \hat{B}_l$. 
			
			$\ket{a_{n,m}}\otimes\ket{b_{n,m}} $ cannot be the eigenstate of composite operators other than the members of of~$\{\compop{A}{n}{B}{n}{1},\compop{A}{n}{B}{n}{2},\cdots,\compop{A}{n}{B}{n}{p-1}\}$. There is no more composite operators that~$\ket{a_{n,m}}\otimes\ket{b_{n,m}} $ is one of its eigenstates, which means there are at most~$(p-1)$ mutually commuting composite operators and this contradicts to the assumption.
			\item If~$\hat{A}_n = \hat{A}_m$ and~$k_n = k_m$, then~$\compop{A}{n}{B}{n}{k_n} = \compop{A}{m}{B}{m}{k_m}$ and this contradicts to the assumption.
		\end{enumerate}
	\end{proof}
	
	\section{Correspondence of Assumption 5 in quantum mechanics}
	\lem{} In~$\mathbb{C}^p\otimes\mathbb{C}^p$,~$\forall \hat{A},\hat{B},\hat{C},\hat{D} \in \{\hat{X},\hat{Z},\hat{X}\hat{Z},\hat{X}\hat{Z}^2,\cdots,\hat{X}\hat{Z}^{p-1}\}$, if~$\hat{A}\ne \hat{C},\hat{B}\ne \hat{D}$ then~$\forall m \in \mathbb{F}^*_p$, $\exists!n\in \mathbb{F}^*_p$ such that~$[\hat{A}\otimes\hat{B}^m,\hat{C}\otimes\hat{D}^n]=0$.
	\begin{proof}
		Let~$\hat{A}=\GGenP{X}{Z}{i},\hat{B}=\GGenP{X}{Z}{j},\hat{C}=\GGenP{X}{Z}{k},\hat{D}=\GGenP{X}{Z}{l}$
		\begin{equation}
		\begin{aligned}
		\left[\hat{A}\otimes\hat{B}^m,\hat{C}\otimes\hat{D}^n\right] &=\hat{A}\hat{C}\otimes\hat{B}^m\hat{D}^n-\hat{C}\hat{A}\otimes\hat{D}^n\hat{B}^m \\
		&=\GGenP{X}{Z}{i}\GGenP{X}{Z}{k}\otimes(\GGenP{X}{Z}{j})^m(\GGenP{X}{Z}{l})^n-\GGenP{X}{Z}{k}\GGenP{X}{Z}{i}\otimes(\GGenP{X}{Z}{l})^n(\GGenP{X}{Z}{j})^m
		\end{aligned}
		\end{equation}
		By using Weyl commutation relation, we may obtain the following relation:
		$$\GGenP{X}{Z}{i}\GGenP{X}{Z}{k}=w^{i_2k_1-i_1k_2}_p\GGenP{X}{Z}{k}\GGenP{X}{Z}{i}$$
		$$(\GGenP{X}{Z}{j})^m(\GGenP{X}{Z}{l})^n=w^{mn(j_2l_1-j_1l_2)}_p(\GGenP{X}{Z}{l})^n(\GGenP{X}{Z}{j})^m$$
		In order to let the commutation relation~$[\hat{A}\otimes\hat{B}^m,\hat{C}\otimes\hat{D}^n]=0$ holds, we must have
		$$w^{i_2k_1-i_1k_2}_pw^{mn(j_2l_1-j_1l_2)}_p=1$$
		$$n=\frac{i_1k_2-i_2k_1}{m(j_2l_1-j_1l_2)}$$
		If~$\hat{A}\ne \hat{C},\hat{B}\ne \hat{D}$ then both numerator and denominator cannot be zero. Moreover both numerator and denominator are elements in~$\mathbb{F}^*_p$. This means~$n$ is a unique element in~$\mathbb{F}^*_p$.
	\end{proof}

	\section{Two understandings of information}
	
	We'd like to emphasize there are two different understanding/interpretations of information, even if they are both function of probability distribution. The forms of probability distributions are different, as well as the function that depend on them.
	
	\subsection{Information as a measure of uncertainty of the outcomes of a measurement}
	In this understanding, the probability distribution of a measurement denotes the probability of its possible outcomes. An experimenter may have already done a sequel of measurements on a same system, say:$$h_{\le t_n}=\{``\hat{A}_1,\lambda_1,t_1",``\hat{A}_2,\lambda_2,t_2",``\hat{A}_3,\lambda_3,t_3",\cdots, ``\hat{A}_n,\lambda_n,t_n"\}$$
	Based on this background knowledge we could~\textit{derive} the probability of the outcome of some other measurements, say probability of measurement~$\hat{B}$ with outcome~$\lambda_B$:
	$$P(``\hat{B},\lambda_B,t_{n+1}"|h_{\le t_n},I)$$
	
	We may express the information of~$\hat{B}$ as~$I(\hat{B}|h_{\le t_n})$ under this background as a function of the outcome probabilities~$\{P(``\hat{B},\lambda_B,t_{n+1}"|h_{\le t_n},I)\}$.
	Several properties are observed of this~$I(\hat{B}|h_{\le t_n})$:
	\begin{enumerate}
		\item \textbf{The measure is similar to Shannon entropy.} Though we didn't put an expression here, this measure describes the degree of uncertainty of the possible outcomes of measurement~$\hat{B}$. It achieves an upper bound which is usually denoted as 1 unit when the probability distribution is a peak distribution and has a lower bound for which the parameter is a uniform probability distribution.
		
		The reason of not using Shannon entropy directly is that when trying to use the information of measurement to compose another measure of information of the system, the Shannon entropy may lead to the violation of unitary transformation invariant\cite{inadequacyShannon}.
		\item \textbf{The probability distribution~$P(``\hat{B},{\lambda},t_{n+1}"|bkgd,I)|_\lambda $ is exact, and it is determined by the mathematical structure of the system.} This probability is actually derived from the background, it needs us to have a fully understanding of the system, as well as the relation between~$\hat{B}$ and the background. For example, for spin-$\frac{1}{2}$ system, we could directly write the following two probabilities:
		$$P(``\hat{\sigma}_x, -1, t_2"|``\hat{\sigma}_x, +1, t_1",I) = 0 $$
		$$P(``\hat{\sigma}_z, +1, t_2"|``\hat{\sigma}_x, +1, t_1",I) = \frac{1}{2} $$
		We could do so based on the knowledge that the state of the system after the measurement of~$\hat{\sigma}_x$ will be remained as an eigenstate of~$\hat{\sigma}_x$ and~$\hat{\sigma}_x$ and~$\hat{\sigma}_z$ are mutually complementary.
	\end{enumerate}
	
	\subsection{Information as a measure of information gain}
	The second understanding mainly related with quantum state tomography, where the probability of the outcome of a specific measurement is obtained through the statistics of measurements on an ensemble of many identical copies of the system. In fact what we deal with is a posterior updated with given data.
	
	For example, we take a set of measurements~$\hat{C}$ on an ensemble of identical systems, and obtain a data sequence~$D_n$. By using Bayesian rule we could get the posterior of measurement~$\hat{C}$:
	$$ P(``\hat{C},\lambda_c"|D_n, I) = \frac{P(D_n|``\hat{C},{\lambda}_c", I)P(``\hat{C},{\lambda}_c"| I)}{P(D_n|I)} $$
	
	The information gain of~$\hat{C}$ through this data~$D_n$ can then be defined using Kullback-Leibler divergence:
	$$I_{KL}(\hat{C}|D_n) = D_{KL} (P(``\hat{C},{\lambda}_c"|D_n, I)||P(``\hat{C},{\lambda}_c"| I)) = \int P(\vec{x}|D_n, I) \log\frac{P(\vec{x}|D_n, I)}{P(\vec{x}|I)} d\vec{x}$$
	where~$P(\vec{x}|D_n, I)$ is the probability density of~$P(``\hat{C},{\lambda}_c"|D_n, I)$, if measurement~$\hat{C}$ has~$n$ different outcomes, then~$\vec{x}$ is a tuple of~$n-1$ parameters.
	
	Several properties observed of~$I_{KL}(\hat{C}|D_n)$:
	\begin{enumerate}
		\item \textbf{The measure is more like to be the information of the posterior under the given data rather than the information of the measurement~$\hat{C}$.} The Kullback-Leibler divergence describes the deviation between the posterior and prior when the prior is fixed.
		\item \textbf{The posterior is obtained directly from the given data, we needn't much knowledge of the mathematical background.} With more and more data, the updated posterior will be more and more close to the probability of outcomes of~$\hat{C}$ of the pre-tuned state. However, the posterior would never be exactly the same with the pre-tuned probability.
	\end{enumerate}

\end{document}